\documentclass[5p]{elsarticle}

\usepackage{lineno,hyperref}

\usepackage{amsmath}
\usepackage{amssymb}
\usepackage{makecell}
\usepackage{array}
\newcolumntype{?}{!{\vrule width 1pt}}

\hypersetup{colorlinks,allcolors=black}
\journal{The European Physical Journal A}

\begin{document}

\begin{frontmatter}

\title{Study of radial motion phase advance during motion excitations in a Penning trap and accuracy of JYFLTRAP mass spectrometer}

\author[a]{D.A.~Nesterenko\corref{mycorrespondingauthor}}
\cortext[mycorrespondingauthor]{Corresponding author}
\ead{dmitrii.nesterenko@jyu.fi}

\author[a]{T.~Eronen}
\author[a]{Z.~Ge}
\author[a]{A.~Kankainen}
\author[a]{M.~Vilen\fnref{fn1}}
\fntext[fn1]{\textit{Current address:} Experimental Physics Department, CERN, CH-1211 Geneva 23, Switzerland}

\address[a]{University of Jyv\"askyl\"a, P.O. Box 35, FI-40014 University of Jyv\"askyl\"a, Finland}

\begin{abstract}
Phase-imaging ion-cyclotron-resonance technique has been implemented at the Penning-trap mass spectrometer JYFLTRAP and is routinely employed for mass measurements of stable and short-lived nuclides produced at IGISOL facility. Systematic uncertainties that impose limitations on the accuracy of measurements are discussed. It was found out that the phase evolution of the radial motion of ions in a Penning trap during the application of radio-frequency fields leads to a systematic cyclotron frequency shift when more than one ion species is present in the trap during the cyclotron frequency measurement. An analytic expression was derived to correctly account for the shift.
Cross-reference mass measurements with carbon-cluster ions have been performed providing the mass-dependent and residual uncertainties.
\end{abstract}

\begin{keyword}
\texttt Mass spectrometers \sep Binding energies and masses \sep Penning trap \sep Carbon clusters
\end{keyword}

\end{frontmatter}


\section{Introduction}

Penning-trap mass spectrometry is a widely used method for very accurate atomic mass measurements, applicable both for stable and radioactive isotopes down to short half-lives ($T_{1/2} \approx $ 100~ms). The phase-imaging ion-cyclotron-resonance (PI-ICR) technique \cite{Eliseev2013} has become increasingly employed in the Penning-trap mass spectrometry.  It provides a higher accuracy, sensitivity and resolving power than the conventional time-of-flight ion-cyclotron-resonance (TOF-ICR) technique \cite{Graff1980,Konig1995}. Higher accuracy is required e.g. for neutrino studies, where accurate mass differences are needed \cite{Nesterenko2014,Eliseev2015}. Due to the sensitivity of the PI-ICR method, more exotic nuclei with low production rates far from the stability can be explored. The superior resolving power of the PI-ICR technique enables the studies of low-lying isomeric states with excitation energies of a few tens of keV, unavailable with other mass-spectrometry methods.

Systematic effects specific to the PI-ICR technique were considered in detail in \cite{Eliseev2014}. These include collisions of the stored ions with residual gas in a Penning trap, the temporal instability of the trapping voltage, the imperfections of the trapping potential and the uncertainty due to the conversion of the cyclotron motion to the magnetron motion limiting the maximum accuracy and resolving power. In addition to these effects, the effects due to the ion-ion interactions and the mass-dependent motional phase advance prior to the phase accumulation time \cite{Orford2020} have also to be taken into account when more than a single ion species is simultaneously stored in the trap. Ions with different masses gain a phase difference, which is significant only when more than one ion species is present. An analytic expression was derived to correct for this shift. 
In this work, we study the various systematic uncertainties in the mass measurements done with the PI-ICR technique at the JYFLTRAP double Penning trap mass spectrometer.
The harmonization procedure of the trapping potential at JYFLTRAP has already been described in \cite{Nesterenko2018}.

Cross-reference mass measurements with carbon-cluster ions provide a way to determine systematic uncertainties for a broader range of masses and cyclotron frequencies, i.e., when the ion of interest and the reference ion have different mass numbers. The carbon clusters have been successfully employed to demonstrate the Penning trap performance at ISOLTRAP \cite{Kellerbauer2003}, SHIPTRAP \cite{Chaudhuri2007},  JYFLTRAP \cite{Elomaa2009} and LEBIT \cite{Bustabad2014} but these have been performed using the TOF-ICR technique. Here we report on the carbon-cluster measurements with the PI-ICR technique at\\ JYFLTRAP.

\section{Experimental setup and method}

\subsection{JYFLTRAP double Penning trap mass spectrometer}

\begin{figure*}[htb]
\centering
\includegraphics[width=0.95\textwidth]{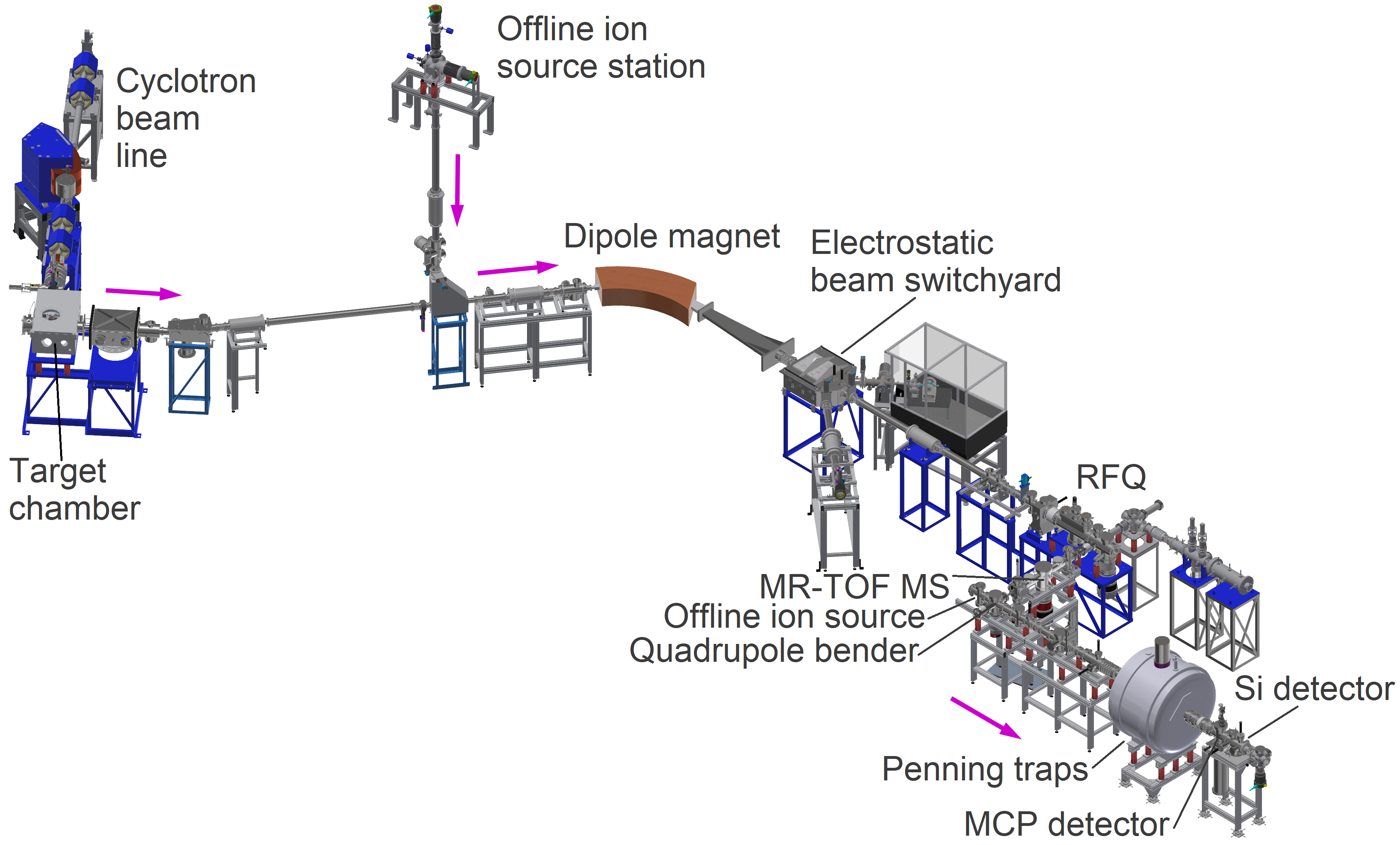}
\caption{Overview of the IGISOL facility, consisting of the target chamber, the 55$^{\circ}$ dipole magnet, the electrostatic beam switchyard, the gas-filled RFQ, the MR-TOF mass spectrometer and the Penning-trap setup JYFLTRAP.
The offline ion source station upstairs and the offline ion source mounted at the quadrupole bender after MR-TOF MS can provide the stable ions for calibration and mass measurements with the Penning traps.}
\label{fig:IGISOL}
\end{figure*}

JYFLTRAP is a double Penning-trap mass spectrometer used for high-precision mass measurements of stable and radioactive ions and high-resolution ion beam purification \cite{Eronen2012} at the Ion Guide Isotope Separator On-Line (IGISOL) facility \cite{Moore2013} (Fig.~\ref{fig:IGISOL}). Radioactive ions are produced via fission, fusion-evaporation or multi-nucleon transfer reactions at IGISOL, stopped in helium gas and extracted via a sextupole ion guide (SPIG) \cite{Karvonen2008} to high vacuum, where they are electrostatically accelerated to 30$q$~kV energy ($q$ is the charge of ions). The ion beam is mass-separated using a 55$^{\circ}$ dipole magnet with a mass resolving power ($R = m/\Delta m$) of about 500. 
The ions with the selected mass-to-charge ratio $A/q$ are electrostatically decelerated to $\sim100$~eV and injected into a gas-filled radio-frequency quadrupole (RFQ) \cite{Nieminen2001}, where they are cooled and bunched. The RFQ is placed at the same +30 kV high voltage platform as the Penning traps.
The bunched ion beam can be additionally purified after the RFQ with the Multi-Reflection Time-of-Flight (MR-TOF) Mass Separator/Spectrometer before injection into JYFLTRAP.

Alternatively, ions of stable isotopes can be produced in the glow discharge, surface ionization or laser ablation ion sources placed in the IGISOL target chamber \cite{Rahaman2008}, at the offline ion source station \cite{Vilen2020}, in front of the RFQ or in front of the Penning traps (discussed below). Typically, the singly charged ions constitute the main fraction of ions produced at IGISOL.

The JYFLTRAP cylindrical double Penning trap system is placed inside a 7-T superconducting magnet (Magnex Scientific). The ions injected into the first (preparation) trap are cooled, centered and purified via a mass-selective buffer gas cooling technique \cite{Savard1991}. This technique allows to separate individual isobars ($R \leq 10^5$) and prepare isobarically pure ion samples for high-precision mass measurements or post-trap decay spectroscopy. The ions are transferred through a 1.5 mm diameter diaphragm between the traps into the second (measurement) trap. There, the mass of the ion with mass $m$ and charge $q$ is determined based on its cyclotron frequency
\begin{equation} \label{eq:qbm}
\nu_{c} = \frac{1}{2\pi}\frac{q}{m}B,
\end{equation}
in the magnetic field $B$.
A position-sensitive microchannel plate (MCP) detector with a delay-line anode (RoentDek GmbH, model DLD40) is located outside the strong magnetic field at ground potential. Thus, the ions extracted from JYFLTRAP hit the detector with $30q$~kV of beam energy. In 2020, the detector was moved further away from the magnet by about of 12 cm compared to the original configuration \cite{Nesterenko2018}. The detector is now about 104 cm from the center of the measurement trap. The detection efficiency is discussed in Section~\ref{sec:efficiency}.

\begin{figure}[htb]
\includegraphics[width=0.49\textwidth]{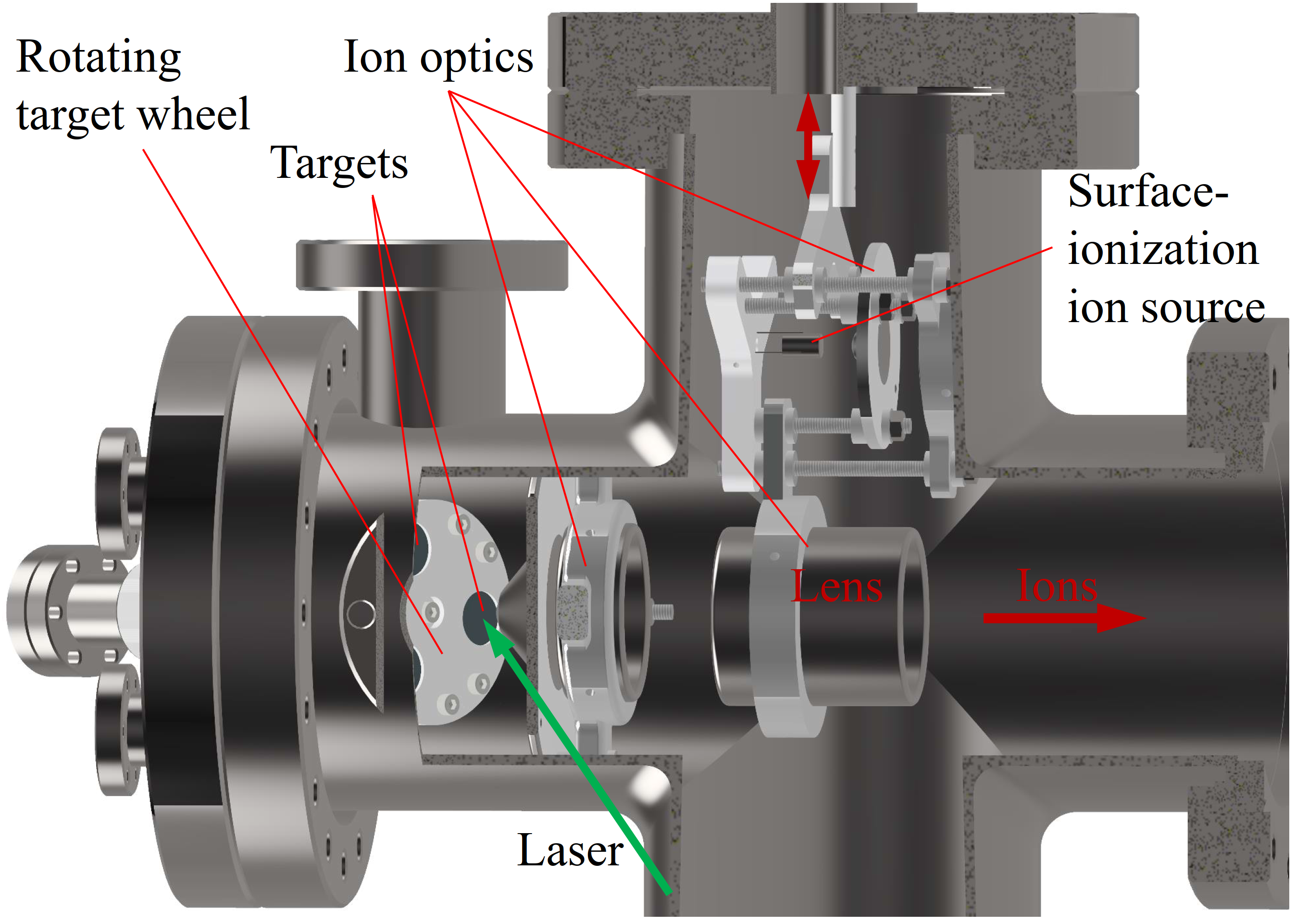}
\caption{Cut view of the offline ions source, showing the laser ablation source in use. The surface-ionization source is mounted on an actuator, which allows either having the source or a pass-through lens for the ions from the laser ablation source to be on the horizontal beam axis.}
\label{fig:ion_source}
\end{figure}

\subsection{Offline ion source for JYFLTRAP \label{sec:ion_source}}

A new offline ion source consisting of a laser ablation and surface ionization ion source (Fig.~\ref{fig:ion_source}) was mounted at the beam line between the MR-TOF mass separator and the Penning traps, see Fig.~\ref{fig:IGISOL}. A quadrupole ion beam bender, located in front of the offline ion source, can transport ions to JYFLTRAP either from the offline ion source (straight direction) or from the RFQ (deflection by 90$^{\circ}$). The potential of the offline ion source is set 50 V lower than the potentials of the trap endcap electrodes to allow ions to be trapped by switching the voltage on the endcaps. The ions from the ion source have 800$q$~V of energy. The position of the actuator allows to choose ions either from the laser ablation ion source (actuator at lens position) or from the surface ionization source (actuator at surface ionization ion source position). Both ion sources utilize a skimmer electrode and electrostatic lens to form an ion beam.

The surface ionization source provides ions of stable isotopes $^{39,41}$K, $^{85,87}$Rb and $^{133}$Cs. They are embedded into the same heatable pellet. The current heating the filament is adjusted based on the required beam intensity and it is typically 1.2 - 1.5 A.

The laser ablation ion source consists of a rotating target holder and Nd:YAG 535 nm laser which operates with up to 10 Hz repetition rate. The laser with the energy per pulse of $>$2 mJ, bombarding the target with a diameter of 16 mm, is focused to a spot size of about 1 mm in diameter. The target holder rotates changing direction within the angles that cover an individual target in order to prevent the laser continuously ablating the same spot. At its current configuration, three targets can be simultaneously installed in the laser ablation ion source. 

\subsection{Penning trap mass spectrometry}

\subsubsection{Basic principle of a Penning trap}

A Penning trap allows storing and manipulation of charged particles in the confining volume of the trap.
At JYFLTRAP, predominantly singly charged ions are trapped for mass measurement or separation.
An ion is confined in a Penning trap by the superposition of a strong homogeneous magnetic field $\vec{B}=B_z\vec{e}_z$ and a quadrupolar electrostatic potential, which has in cylindrical coordinates $(z,\rho)$ the form
\begin{equation} \label{eq:quadrupolar_potential}
V(z, \rho) = \frac{U_0}{2d^2}(z^2-\rho^2/2),
\end{equation}
where $z$ and $\rho=\sqrt{x^2+y^2}$ are the axial and radial distance from the trap center, respectively, $U_0$ is the trap potential (the potential difference between the ring and the endcap electrodes \cite{Eronen2012}) and $d$ is the characteristic dimension of the trap defined by the trap geometry. 

The trajectory of an ion in the trap is a superposition of three independent, ideally harmonic, eigenmotions. One of the motions, called the axial motion, occurs along the magnetic field lines at axial frequency $\nu_z$. The other two motions are radial motions perpendicular to the magnetic field with the frequencies $\nu_-$ and $\nu_+$. The magnetron motion with the smaller $\nu_-$ magnetron frequency is almost mass independent, while the cyclotron motion described by the modified cyclotron frequency $\nu_+$ is mass dependent. 
The frequencies of eigenmotions are expressed as \cite{Brown1986}:
\begin{equation} \label{eq:axial_freq}
\nu_{z} = \frac{1}{2\pi} \sqrt{\frac{q U_0}{m d^2}},
\end{equation}
\begin{equation} \label{eq:radial_freq}
\nu_{\pm} = \frac{1}{2} (\nu_{c} \pm \sqrt{\nu_{c}^2-2\nu_{z}^2}),
\end{equation}
and the frequency hierarchy is $\nu_c\approx\nu_+\gg\nu_z\gg\nu_-$.

The sum of the two radial motion frequencies is the ion cyclotron frequency $\nu_c$:
\begin{equation} \label{eq:sum_of_radial_frequencies}
\nu_c = \nu_+ + \nu_-.
\end{equation}
This relation is valid in case of an ideal Penning trap, where the electric potential is fully harmonic, the magnetic field is perfectly homogeneous and there is no misalignment between the magnetic and electric field axis. A more robust relationship, called the invariance theorem \cite{Brown1982,Gabrielse2009}, is fulfilled in the case of a real Penning trap:
\begin{equation} \label{invariance_theorem}
\nu_c^2 = \nu_+^2 + \nu_-^2 + \nu_z^2.
\end{equation}

The ion is manipulated in a Penning trap by applying radiofrequency (rf) fields in different configurations. A dipolar rf excitation at an eigenfrequency of the ion ($\nu_+$, $\nu_-$, $\nu_z$) can be used to excite the corresponding ion motion in the trap. A quadrupolar rf field at the cyclotron frequency $\nu_c$ allows to convert one radial motion to the other \cite{Konig1995}. The ring electrodes of the traps at JYFLTRAP are eight-fold segmented \cite{Eronen2012}, allowing the application of the dipolar, quadrupolar and octupolar rf fields in the radial plane. For the dipolar excitation of the axial motion the voltages are applied to the endcap electrodes.

\subsubsection{Cyclotron frequency determination with Phase-Imaging Ion-Cyclotron Resonance technique\label{sec:PI-ICR}}
The PI-ICR method  \cite{Eliseev2013,Eliseev2014} is based on observation of phase evolution of radial ion  motions  in  a Penning  trap. The radial position of ions in the trap is projected onto a position-sensitive MCP detector with a certain magnification factor. Ideally, the projection preserves the angular relations magnifying the relative positions of the ions on the detector. The PI-ICR method allows to determine the radial frequencies independently or measure directly their sum, i. e. the cyclotron frequency $\nu_c$ (Eq.~(\ref{eq:sum_of_radial_frequencies})). 

For the direct cyclotron frequency determination two excitation patterns, differing by only one step, are applied alternately (Fig.~\ref{fig:scheme}). First, the ions that have been cooled, centered and purified in the preparation trap, are transferred to the center of the measurement trap (step 1). Then, the coherent components of the magnetron and axial motions are reduced by dipolar rf pulses at the corresponding frequencies (steps 2a and 2b). After these preparatory steps the cyclotron motion of the ions is excited via a dipolar rf pulse at the frequency $\nu_+$ (step 3),
imprinting a cyclotron phase. 
The step 4 is the one that differs in the two excitation patterns. This step is a qudrupolar rf pulse at the cyclotron frequency $\nu_c$, converting the cyclotron motion to the magnetron motion.
In pattern 1, the conversion $\nu_c$-pulse is applied immediately after the $\nu_+$-pulse. After, the ions rotate freely and the magnetron motion accumulates a magnetron-motion phase $\phi_- + 2 \pi n_- = 2 \pi \nu_- t_{acc}$ during phase accumulation time $t_{acc}$, where $\varphi_- \in [0,2\pi)$ is the phase of the last orbital period and $n_-$ the integer number of revolutions. Subsequently, the ions are ejected from the trap (step 5) and their position is projected onto the detector. Position of the ion image of the magnetron phase on the detector is described by the polar angle $\alpha_-$ with respect to the trap center (projection of ions from the center of the trap).
In pattern 2, the quadrupolar rf pulse at the cyclotron frequency $\nu_c$ is applied after the time $t_{acc}$, allowing ion motion to accumulate a cyclotron-motion phase $\phi_+ + 2 \pi n_+ = 2 \pi \nu_+ t_{acc}$, similarly to the magnetron-motion phase, where $\varphi_+ \in [0,2\pi)$ and $n_+$ the integer number of revolutions. After conversion, the ions are  projected onto the detector giving the image of the cyclotron phase at the polar angle $\alpha_+$ with respect to the trap center. Note, that the conversion preserves the modulus of the angle of the accumulated cyclotron phase and flips the sign of the angle (see Sec.~\ref{sec:excitations}). Thus, the ion cyclotron motion and the movement of the image on the detector have opposite angular directions.

\begin{figure}[htb]
\includegraphics[width=0.49\textwidth]{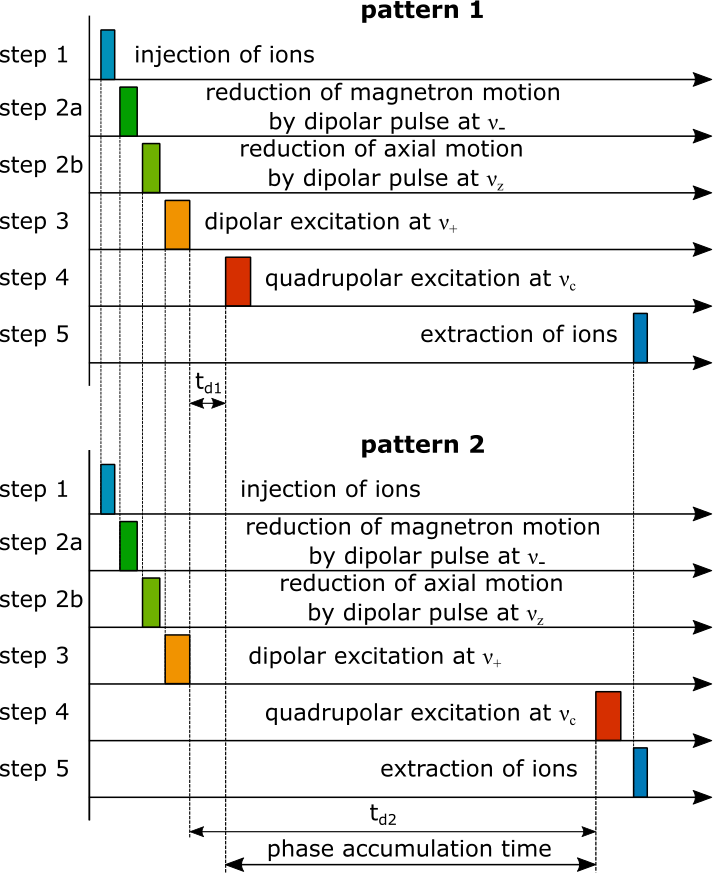}
\caption{Measurement sequence in the measurement trap for the cyclotron frequency ($\nu_c$) determination with the PI-ICR technique. The magnetron and cyclotron phases are accumulated in the patterns 1 and 2, respectively, differing by the position of the conversion pulse of the quadrupolar excitation.}
\label{fig:excitation_scheme}
\end{figure}

We emphasize that it is the difference in time of conversion pulse at step 4 between pattern 1 and 2, i.e., the phase accumulation time $t_{acc}$, that is critical fo $\nu_c$ measurement. It needs to be known with a sub-ns precision. With the exception of the timing of step 4, the two patterns are identical.
The cyclotron frequency is determined as
\begin{equation} \label{eq:freq_alpha}
\nu_{c}= \nu_{-} + \nu_{+} = \frac {\alpha_c + 2 \pi n_c} {2 \pi t_{acc}},
\end{equation}
where $\alpha_c = \alpha_+ - \alpha_-$ is the angle between the two phase images, $n_c$ is the full number of revolutions, which the studied ions would perform in a magnetic field $B$ in absence of electric field during a phase accumulation time $t_{acc}$. For initial unambiguous $n_c$ determination, the cyclotron frequency is determined, e.g., by a quick TOF-ICR measurement with a moderate precision. 
The number $n_c$ can also be deduced with few preliminary PI-ICR measurements starting with short phase accumulation time to keep track of $n_c$.
It is enough to perform such a measurement only once every several days due to a rather small drift of the magnetic field (see Sec.~\ref{sec:magnet_stability}). 

Precision of the cyclotron frequency determination depends on the duration of the phase accumulation time $t_{acc}$, the angular sizes of the phase spots and the number of detected ions $N$. To understand the uncertainty of the frequency determination let us assume the magnetron and cyclotron phase spots are circular, have the same size, the same radial distance to the trap center ($r_- \cong r_+ \equiv r$) and the same number of detected ions. Therefore, the statistical uncertainty of the cyclotron frequency can be written as:
\begin{equation} \label{eq:freq_uncertainty}
\delta \nu_{c} = \frac{\delta \alpha_{c}}{2 \pi t_{acc}} = \frac{\Delta r}{\pi r t_{acc} \sqrt{N}},
\end{equation}
where $\Delta r$ is the standard deviation of ion distribution of the magnetron or cyclotron phase spot on the detector and $N$ is the total number of detected ions ($N/2$ for individual mangetron and cyclotron spots) and assuming no position uncertainty for the center.
Initially, the ion distribution is defined by the cooling in the preparation trap. The instability of trapping potential and ion collisions with atoms of residual gas in the measurement trap are the main effects that change the radial motions and result in an increase in the spread of the phase spots on the detector \cite{Eliseev2014}.
Both of these effects are stronger with increasing phase accumulation time $t_{acc}$, thus, limiting $t_{acc}$ typically to $\le 1.2$~s at JYFLTRAP.

\subsubsection{Radiofrequency excitations\label{sec:excitations}}

The duration of the excitation pulses can be chosen as short as one period of the corresponding rf excitation frequency. In practice, the minimum duration is limited by the amplitude of the pulse that can be applied from the function generators
and typically is several periods in order to gain the desired effect of the excitation. 

The dipolar excitation at the modified cyclotron frequency $\nu_{+}$ is applied in the measurement scheme to excite the ions (Fig.~\ref{fig:excitation_scheme}, step 3). The dipolar excitation is formed by applying two voltages with the phase shifted by $\pi$ to two opposite segments of the ring electrode. The potential created by the dipolar excitation in a Penning trap is 
\begin{equation} \label{eq:dipolar_potential}
V_d(t) = \frac{a U_d x}{\rho_0} \cos(2 \pi \nu_d t + \phi_d),
\end{equation}
where $U_d$, $\nu_d$ and $\phi_d$ are the amplitude, frequency and the initial phase of the dipolar excitation, $\rho_0$ is the inner radius of the ring electrode and $a$ is a geometry factor for the trap electrodes \cite{Blaum2003}. The potential $V_d(t)$ is added to the trapping potential $V(z, \rho)$ (Eq.~\ref{eq:quadrupolar_potential}) of the trap.

The equations for the radial motion in the trap can be solved by introducing the velocity vectors \cite{Brown1986}:
\begin{equation} \label{eq:velocity_vectors}
\vec{V}^{\pm} = \dot{\vec{\rho}} - 2 \pi \nu_{\mp} \vec{\rho} \times \hat{z},
\end{equation}
where $\vec{\rho} = (x, y, 0)$. Following the procedure described in \cite{Blaum2003} the solution of the equations for $\vec{V}^{\pm}$ can be found using the ansatz
\begin{equation} \label{eq:anzatz}
\vec{V}^{\pm}(t) = \vec{A}^{\pm}(t)e^{\pm i(2\pi \nu_\pm t + \phi_\pm^0)}.
\end{equation}
Resonant cases when the frequency $\nu_d$ is close to the one of the radial eigenfrequencies, either $\nu_+$ or $\nu_-$, are considered. The excitation at the frequency $\nu_d \approx \nu_+$ does not affect the amplitude $\vec{A}^{-}$ of the magnetron motion and, similarly, the excitation at $\nu_d \approx \nu_-$ does not affect the amplitude $\vec{A}^{+}$ of the cyclotron motion.
It is assumed that the ion motion in the trap remains circular, i.e. $A^{\pm}_x = \mp i A^{\pm}_y$ and the equations for the $x$ and $y$ components are combined. Marking the solution for the amplitude near the resonant modified cyclotron frequency and magnetron frequency of excitation with index ``+'' and ``-'', respectively, it can be written as \footnote{It should be noted that the solution shown here is slightly different than the one given in \cite{Blaum2003}.}
\begin{equation} \label{eq:Ad_solution}
A^{\pm}_y(t) = A^{\pm}_y(0) + \frac{k}{4 \pi \Delta \nu_{\pm} } e^{\pm i \Delta \phi_\pm} (1 - e^{\pm i 2 \pi \Delta \nu_{\pm} t}),
\end{equation}
where $k = q a U_d / (m \rho_0)$, $\Delta \nu_{\pm} = \nu_d - \nu_{\pm}$ is the frequency detuning parameter, $\Delta \phi_\pm = \phi_d - \phi_\pm^0$ is the phase difference of the rf field and the existing ion motion, $A^{\pm}_y(0) = 2 \pi (\nu_+ - \nu_-) \rho_{\pm}(0)$ and $\rho_{\pm}(0)$ are the initial radii of the cyclotron motion and magnetron motion. 

The component of the complex velocity vector $V^{\pm}_y(t) = |V^{\pm}_y(t)|e^{\pm i \phi_{\pm}(t)}$ describes the radial motion, where $\phi_{\pm}(t)$ is the phase of the corresponding radial motion. The radii of the radial eigenmotions are given by:
\begin{multline} \label{eq:dipolar_radii}
\rho_{\pm}(t) = \frac{|V^{\pm}_y(t)|}{2 \pi (\nu_+ - \nu_-)} = \frac{1}{2 \pi (\nu_+ - \nu_-)} \left[ A^{\pm 2}_y(0) + \right.\\
\left.+ \frac{k^2}{(2 \pi \Delta \nu_{\pm})^2} \sin^2(\pi \Delta \nu_\pm t) + \right.\\
\left.+ \frac{A^{\pm}_y(0) k}{2 \pi \Delta \nu_{\pm}} \left( \cos(\Delta \phi_\pm) - \cos(2 \pi \Delta \nu_{\pm} t + \Delta \phi_\pm) \right) \right] ^{1/2} \xrightarrow[\Delta \nu_{\pm} \rightarrow 0]{} \\ \frac{1}{2 \pi (\nu_+ - \nu_-)} \left[ A^{\pm 2}_y(0) + \frac{k^2 t^2}{4} + A^{\pm}_y(0) k t \sin(\Delta \phi_{\pm}) \right] ^{1/2}
\end{multline}

Let us consider the situation when the ions injected into the center of the measurement trap have a zero radius of the cyclotron motion, i.e. $A^{+}_y(0) = \rho_{+}(0) = 0$. Then, the dipolar excitation at the frequency $\nu_d \approx \nu_+$ is applied at the time $t = 0$. The amplitude $|V^{+}_y(t)|$ and the phase $\phi_{+}(t)$ of the cyclotron motion over the excitation time are given by:
\begin{multline} \label{eq:dipolar_exc_motion}
|V^{+}_y(t)|e^{ i \phi_{+}(t)} = | \frac{k}{2 \pi \Delta \nu_{+}} \sin(\pi \Delta \nu_{+} t)| \cdot \\ \cdot e^{i (2 \phi_d + 2\pi (\nu_d + \nu_+) t + 3\pi)/2} \xrightarrow[\Delta \nu_{+} \rightarrow 0]{} \\
\frac{k t}{2} e^{i (2 \phi_d + 2\pi (\nu_d + \nu_+) t + 3\pi)/2}.
\end{multline}
Hence, the phase is
\begin{equation} \label{eq:dipolar_exc_phase}
\phi_{+}(t) = \phi_d + \pi (\nu_d + \nu_+) t + 3\pi/2 + 2 \pi p,
\end{equation}
where $p$ is an integer. The phase evolution in time depends on both $\nu_d$ and $\nu_+$ frequencies, and when they are equal, it changes as $\phi_{+}(t) \propto 2 \pi \nu_+ t$, equally to that of the free cyclotron rotation in absence of the dipolar rf field. The cyclotron phase accumulated during dipolar excitation is the same in both excitation patterns (step 3 in Fig.~\ref{fig:excitation_scheme}) and its effect is cancelled out. The effect of the finite duration of the dipolar excitation at $\nu_+$ frequency, when more than one ion species is in the measurement trap, is discussed in Section~\ref{sec:magnetron_phase}.

Let us consider now the quadrupolar excitation of the conversion pulse that is achieved by the application of the rf voltage to four segments of the ring electrode at the frequency close to the cyclotron frequency $\nu_c$, with the same phase to the opposite segments and with the phase shifted by $\pi$ to the neighboring segments. The theoretical description of the conversion of ion's radial motions in a Penning trap by an external rf quadrupolar field is given in \cite{Konig1995, Kretzschmar2012}. The influence of the conversion pulse on the motion phase in PI-ICR measurement is discussed in \cite{Eliseev2014}. In the ideal case the ions perform a pure cyclotron motion with the phase $\phi_+(0)$ at time zero before the conversion. In this case, the conversion pulse initiated at time zero and lasting for a duration $\tau$ converts the cyclotron motion phase of the ions into the magnetron motion phase $\phi_-(\tau)$, which is given by
\begin{equation} \label{eq:conversion_exc_phase}
\phi_{-}(\tau) = \left( 2 \pi \nu_- + \pi (\nu_q - \nu_c) \right) \tau + \phi_q - \phi_+(0) + \frac{3 \pi}{2} + 2 \pi j,
\end{equation}
where $\nu_q$ and $\phi_q$ are the frequency and initial phase of the quadrupolar excitation, and $j$ is an integer. In case of the complete conversion, which can only occur when $\nu_q = \nu_c$ and the excitation amplitude is chosen properly, the radius of the magnetron motion after the conversion is equal to the radius of the cyclotron motion before conversion. 

It can be seen from Eq.~\ref{eq:conversion_exc_phase} that the conversion flips the sign of the initial phase of the cyclotron motion $\phi_+(0)$. The difference of a certain final phase and the reference phase of the cyclotron motion $\phi_+^f - \phi_+^r$ is converted to the phase difference of the final and reference phase of the magnetron motion $\phi_-^f - \phi_-^r$ as follows
\begin{equation} \label{eq:cyc_to_mag_phase}
\phi_-^f - \phi_-^r = -(\phi_+^f - \phi_+^r).
\end{equation}
Thus, the complete conversion preserves the modulus of the angle between the phases and flips the sign of the angle.
A typical duration of the conversion pulse is 2 ms at JYFLTRAP. The excitation patterns for the magnetron and cyclotron phases are applied alternately, i.e. typically every $0.2-2$~s. Daily fluctuations of the magnetron $\nu_-$ and cyclotron $\nu_c$ frequencies at JYFLTRAP (see Sec.~\ref{sec:magnet_stability}) are so small that the errors due to temporal instability of the $\nu_-$ and $\nu_c$ frequencies are negligible in the angle determination.

The error of the angle determination occurs when the ions have a certain magnetron-motion amplitude before the conversion \cite{Eliseev2014}. This error vanishes when 
\begin{equation} \label{eq:cancel_conversion_error}
2 \pi \nu_c t_{acc} = \pi j,
\end{equation}
where $j$ is an integer, i.e. the angle error due to conversion is eliminated when the phase accumulation time $t_{acc}$ is a multiple of half the period of the cyclotron frequency \cite{Kretzschmar2012}. The start times of the conversion pulses at JYFLTRAP are set by the delays $t_{d1}$ and $t_{d2}$ on the function generators for pattern 1 and pattern 2, respectively (Fig.~\ref{fig:excitation_scheme}). The delay times are chosen as the number of periods of applied rf excitation at the frequency $\nu_q \approx \nu_c$, providing the condition of Eq.~(\ref{eq:cancel_conversion_error}) for the phase accumulation time. It can be shown that at $\nu_q = \nu_c$ the magnetron and cyclotron phase spots have the same angular positions on the detector, i.e. $\alpha_- = \alpha_+$, which also reduces the error due to the distortion of ion motion projection (Sec.~\ref{sec:projection_distortion}). Additionally, to eliminate the error due to the residual magnetron motion, the start time of the dipolar excitation pulse at the frequency $\nu_+$ is scanned over the magnetron period to average the phase spot position on the detector over the magnetron phases.

Similarly, to eliminate the shift due to the possible residual cyclotron motion in case the conversion is incomplete, the extraction time from the measurement trap is also scanned over the period of the modified cyclotron frequency. It allows to average the phase spot position on the detector over the phases of the residual cyclotron motion. This scan results in a smearing of the phase spot at the detector by the angle $2 \pi \nu_- / \nu_+ \lesssim 1^{\circ}$, which is practically less than the initial spatial distribution of the ions. Thus, a two-dimensional timing scan over the start time of the dipolar excitation (step 3 in Fig.~\ref{fig:excitation_scheme}) and the extraction time from the trap (step 5 in Fig.~\ref{fig:excitation_scheme}) is carried out at JYFLTRAP during the PI-ICR measurements.

\subsubsection{Calibration of the magnetic field}

The mass of an ion of interest can be derived from the measured cyclotron frequency $\nu_c$ if the magnetic field is known (Eq.~\ref{eq:qbm}). For calibration of the magnetic field, reference ions with precisely-known mass values are used. The mass is determined based on the measured cyclotron frequency ratio $r$ between the reference ion ($\nu_{c, ref}$) and the ion of interest ($\nu_{c, ioi})$:
\begin{equation} \label{eq:freq_ratio}
r = \frac{\nu_{c, ref}}{\nu_{c, ioi}} = \frac{q_{ref}}{q_{ioi}} \frac{m_{ioi}}{m_{ref}},
\end{equation}
where $m_{ioi}$ and $q_{ioi}$ are the mass and charge of the ion of interest, respectively, and $m_{ref}$ and $q_{ref}$ are the mass and charge of the reference ion, respectively. When singly-charged ions are measured, the atomic mass $M_{ioi}$ can be determined as
\begin{equation} \label{eq:atom_mass}
M_{ioi} = (M_{ref} - m_{e}) r + m_{e} + \{r B_{e, ref} - B_{e, ioi} \},
\end{equation}
where $M_{ref}$ is the mass of the reference atom, $m_e$ is an electron mass, $B_{e, ref}$ and $B_{e, ioi}$ are binding energies of valence electron in reference atom and atom of interest, respectively. Typically, the binding energy of a valence electron is less than 10 eV \cite{Lotz1970} and the term in curly bracket can be neglected.

Stable nuclides $^{85}$Rb and $^{133}$Cs with mass uncertainties 5~eV/$c^2$ and 8~eV/$c^2$ \cite{Wang2021}, respectively, are commonly used for the magnetic field calibration at the JYFLTRAP. Carbon clusters $^{12}$C$_n$, employed for the studies of systematic uncertainties in this work, have the following advantages. They cover a broad mass range of the chart of nuclides in the steps of 12 atomic mass units and their masses are very well-known, since the atomic mass unit $u$ is defined as 1/12 of the $^{12}$C mass. 
The molecular binding energy per atom of the clusters are well known and range from 3.1~eV in C$_2$ to 7~eV in C$_{60}$ \cite{Tomanek1991}. The reference ion is selected with the mass as close as possible to the mass of the ion of interest to reduce the systematic uncertainties (Sec.~\ref{sec:clusters}).

In order to determine the cyclotron frequency ratio $r$ the cyclotron frequencies of the reference ion $\nu_{c, ref}$ and ion of interest $\nu_{c, ioi}$ are measured alternately. The measurement time of the cyclotron frequency is defined as the midpoint of time between the start and end of the measurement. To temporally overlap the ion-of-interest and reference ion measurements two calculation procedures have been recently used. In the first well-established (interpolation) method, the frequency $\nu_{c, ref}$, measured before ($t_1$) and after ($t_3$) the measurement time $t_2$ of the frequency $\nu_{c, ioi}$, is linearly interpolated to the time $t_2$ of $\nu_{c, ioi}$ measurement
\begin{equation} \label{eq:interpolation}
\nu^{inter}_{c, ref}(t_2) = \nu_{c, ref}(t_1) + \frac{t_2-t_1}{t_3-t_1} (\nu_{c, ref}(t_3)-\nu_{c, ref}(t_1))
\end{equation}
and a single frequency ratio $r_i= \nu_{c,ref}^{inter}(t_2) / \nu_{c,ioi}(t_2)$ is determined. The final frequency ratio $r$ is a weighted mean ratio of individual ratios $r_i$ with the maximum of internal and external error \cite{Birge1932}. 

In the second (polynomial) method (described, for example, in \cite{Fink2012}) full sets of $\nu_{c, ref}$ and $\nu_{c, int}$ frequencies are simultaneously fitted with $n$-order polynomials $P_n(t)$ and $r \cdot P_n(t)$, respectively, differing only by a coefficient of proportionality $r$. The cyclotron frequency ratio $r$ is one of the fit parameters, the other parameters describe the magnetic field drift and, hence, the cyclotron frequency changes in time. The polynomial order (often $<10$) is chosen to reach the smallest reduced $\chi ^2$. The interpolation and polynomial methods used in our analysis gave mutually agreeing results.

\section{Investigation of the systematic uncertainties at JYFLTRAP}

\subsection{Distortion of the ion motion projection\label{sec:projection_distortion}}

\begin{figure}[h!]
\includegraphics[width=0.49\textwidth]{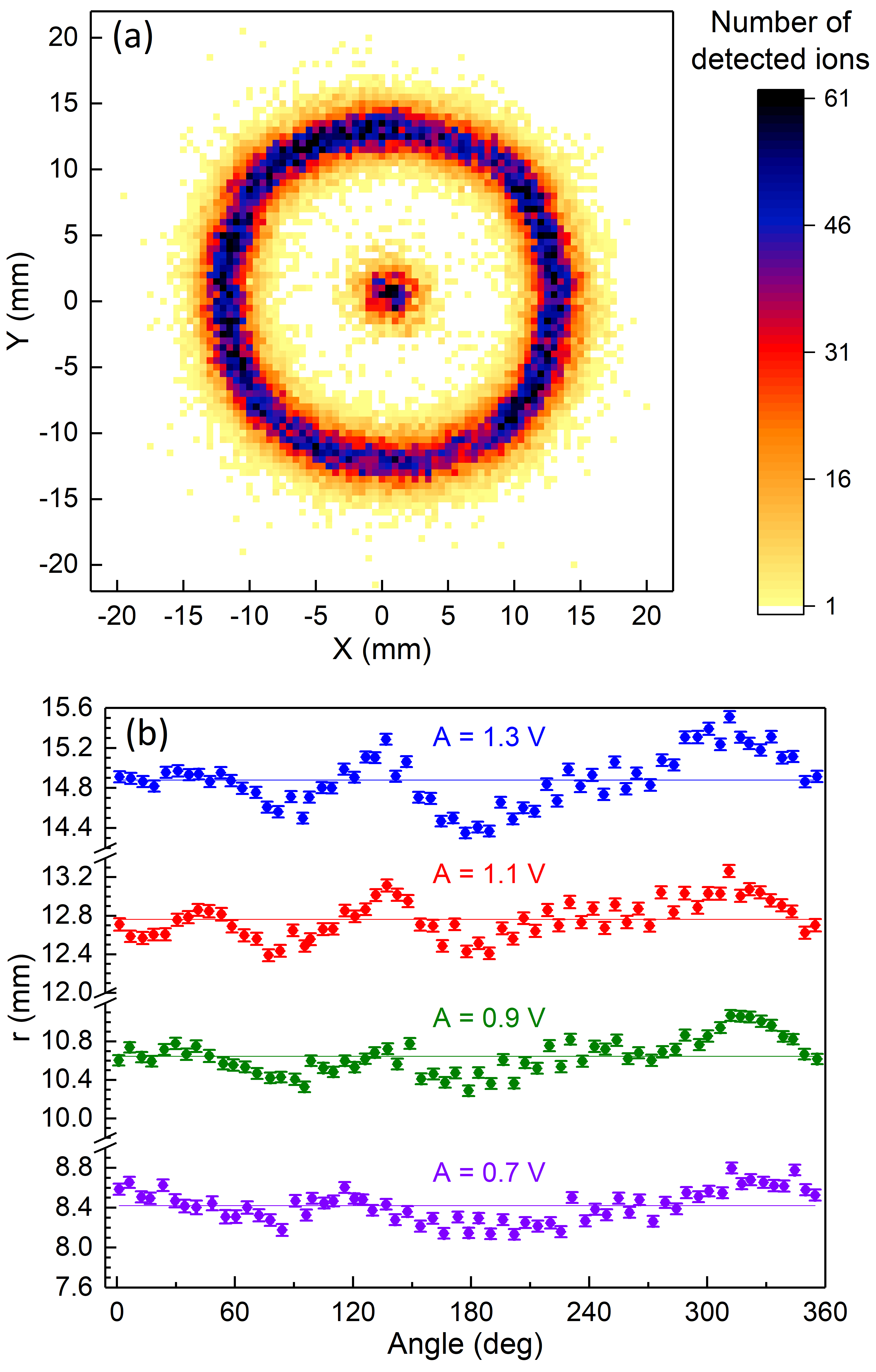}
\caption{(a) Measured projections of $^{133}$Cs$^+$ ions for 63 phases of the magnetron motion and trap center on the detector. The ion motion was excited in the trap by applying a 1-ms dipolar rf pulse at the magnetron frequency $\nu_-$ with amplitude of 1.1 V. (b) Measured radius as a function of the angular position of the phase spot on the detector for different excitation amplitudes. 0$^{\circ}$ angle corresponds to the positive x-axis and angle increases counterclockwise.}
\label{fig:phase-radii}
\end{figure}

\begin{figure}[htb]
\includegraphics[width=0.49\textwidth]{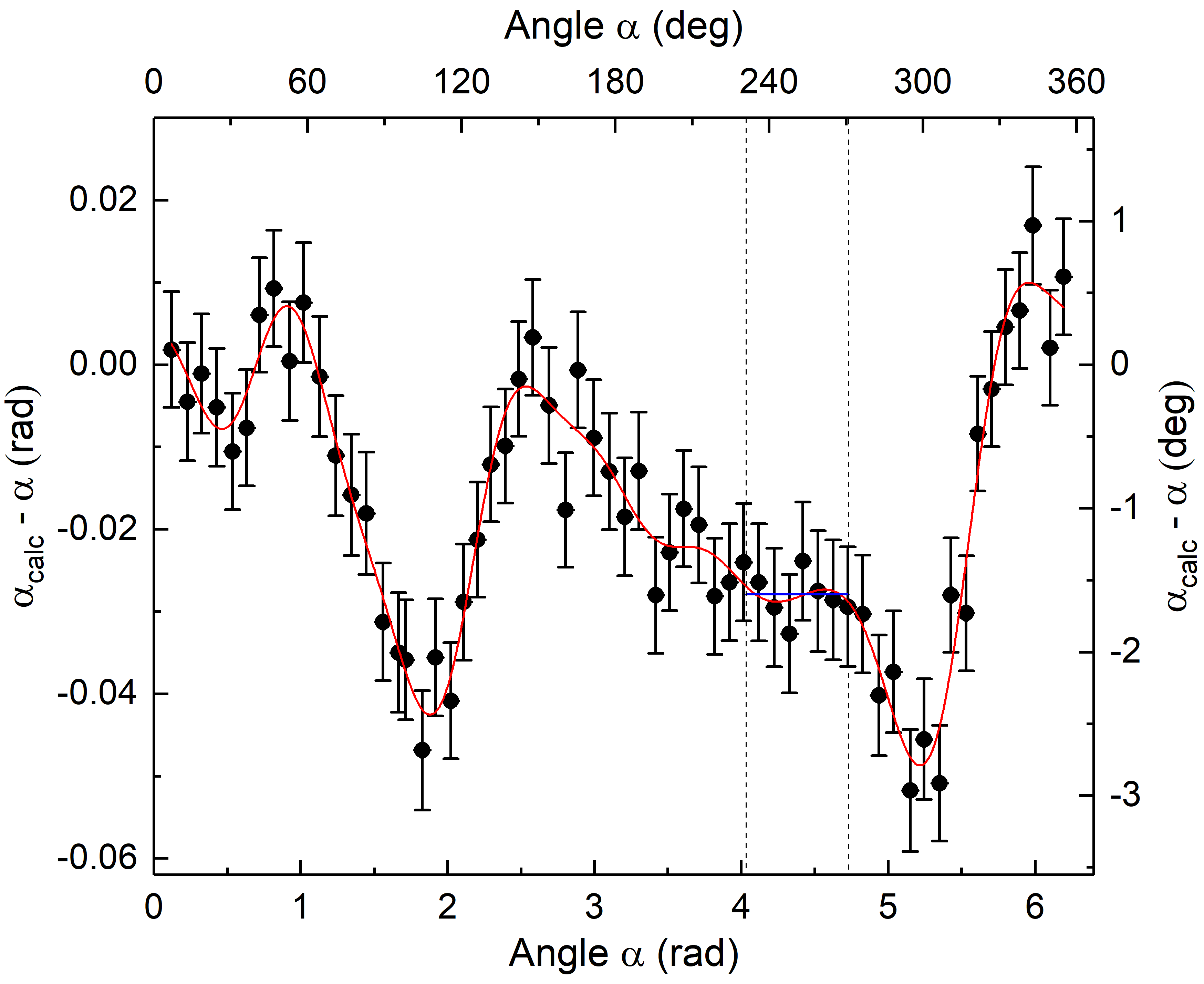}
\caption{Difference between the calculated and measured angles as a function of the angular position of the phase spot on the detector for the excitation amplitude of 1.1 V. The preferable range of angles used in the PI-ICR measurements is indicated by the vertical dashed lines.}
\label{fig:angle-shift}
\end{figure}

Distortion of the ion motion projection onto the detector can be due to misalignment between the magnetic and electric field axes and tilt of the detector plane with respect to the symmetry axis of the trap electrodes. Ions extracted from the measurement trap initially follow the diverging magnetic field lines in the region of constant electric field and, then, accelerated to 30$q$ kV of energy, which creates an intermediate focus point in front of the ground electrode \cite{Nesterenko2018}. Thus, the assumption of the electric field-free drift region \cite{Eliseev2014} is not applicable in the case of JYFLTRAP. After the focus point the beam freely develops and reach the detector.

Distortion of the projection was studied using the magnetron motion, which has a period $T_- \approx 605$~$\mu$s ($\nu_- \approx 1653$ Hz).
The magnetron motion of $^{133}$Cs$^+$ ions was excited in the measurement trap by applying a dipolar rf pulse with a duration of 2 magnetron periods at the frequency $\nu_-$ and amplitude $A_-$ before extracting them with different phases (altogether 63 phase points spaced apart from each other by delaying the extraction of ions from the trap in steps of 9.68~$\mu$s). The start time of the magnetron excitation was scanned over a magnetron period to average possible shifts of the phase spots due to the magnetron motion existing before the excitation. 
The center spot was collected after every two rounds of phase scans. This was done by applying no excitation and by scanning the extraction time from the trap over a magnetron period to average out any residual magnetron motion.
Ideally, the projection would lead to a perfect circle with a certain constant radius, defined as the distance between the phase spot and center spot on the detector. Figure~\ref{fig:phase-radii} shows the measured variation of radius at different angular positions of the phase spots. The radius variations are similar for different amplitudes of the magnetron excitation ($0.7-1.3$~V).

Knowing the magnetron frequency (period) and the time difference between the phases in the measurement, the expected angles between the phase spots were calculated. The reference phase spot position was taken at the angular position 0$^{\circ}$ and deviation of the measured angle between the phase spot and reference phase spot from the calculated value was determined.
This deviation as a function of the angular position of the phase spot on the detector is shown in Figure~\ref{fig:angle-shift} for the amplitude of magnetron excitation of 1.1~V. 
Note, the pulse generator (SpinCore, model PB24-100-4k-PCI), generating the timing triggers for the measurement cycle, has a resolution of 10~ns and the TTL-to-optical converter of the signal introduces the main uncertainty in the timings of 25~ns of jitter. However, this contributes to the uncertainty of the angle determination by an order of magnitude less than the statistical error.
The deviation of the angles for each measurement with a certain amplitude of excitation was fitted with a periodic function $f(\alpha) = A_0 + \sum_{k=1}^{9} A_k \sin(k\alpha - a_k)$, where $A_0, A_k$ and $a_k$ are constants. 
Since the cyclotron frequency is determined via the angle $\alpha_c$, which is the difference of the polar angles $\alpha_+$ and $\alpha_-$ (see Eq.~\ref{eq:freq_alpha}), the uncertainty related to the angle shift depends on the value of $\alpha_{c}$ and the angles $\alpha_+$ and $\alpha_-$ of the phase spot positions on the detector.
To take into account the angle shift in the cyclotron frequency measurement the systematic uncertainty $\delta_{syst}\alpha_{c} = |f(\alpha_+) - f(\alpha_-)|$ can be quadratically added to the statistical uncertainty of the angle $\alpha_{c}$.

Note, that the cyclotron frequency uncertainty caused by the angle shift decreases with increasing the phase-accumulation time $t_{acc}$ (Eq.~\ref{eq:freq_uncertainty}). In precision mass measurements the angle between the cyclotron and magnetron phase spots $\alpha_c$ is tuned to be as close to zero as possible, i. e. the phase-accumulation time is as close to multiple of the period of the cyclotron frequency $\nu_c$. In practice, the angle $\alpha_c$ can remain within a few degrees for several hours with typical fluctuations of the magnetic field (Sec.~\ref{sec:magnet_stability}). The position of the phase spots is also chosen to lie in the region where the angle shift is almost constant, and, thus, the systematic uncertainty for the angle $\alpha_c$ is canceled out. 
For example, such region is in the ranges of polar angles of $231^{\circ}-271^{\circ}$, where $\delta_{syst}\alpha_{c} \leq 0.001$~rad.
This translates to an upper limit for the relative uncertainty of the cyclotron frequency determination of, e.g., singly-charged ions of $^{133}$Cs to about of $2\times 10^{-10}$[s]/$t_{acc}$. This uncertainty is several times smaller than the typical statistical uncertainty. At worst, the magnetron and cyclotron spot images can deviate by 0.05 rad. This translates the relative uncertainty of $1.2 \times 10^{-8}$[s]/$t_{acc}$ and can introduce a significant systematic error. It was observed that the distortion of the projection depends on the preparation of ions in the preparation trap and can differ for different conditions. Thus, the mapping of the optimal angle range can be performed prior to precise PI-ICR measurements in order to minimize the systematic uncertainty of the angle $\alpha_c$.

\subsection{Temporal stability of the magnetic and electric fields\label{sec:magnet_stability}}

Stability of the radial frequencies $\nu_+$ and  $\nu_-$ depend on the stability of the magnetic field $B$ and trapping potential $U_0$ (Eq.~\ref{eq:radial_freq}), while the cyclotron frequency $\nu_c$ depends only on the magnetic field $B$ (Eq.~\ref{eq:qbm}). The magnetic field strength of a superconducting magnet is changing over time. A decrease of the current of the superconducting coils due to the flux creep phenomenon \cite{Anderson1964} leads to a decrease of the magnetic field. This slow change over the years is approximately linear and can be taken into account with calibration. There are also non-linear fluctuations of the magnetic field caused by the instability of the environmental conditions such as temperature, pressure and ambient fields.
The temperature of the IGISOL experimental hall is stabilized at the level below 1~$^{\circ}$C and additional temperature stabilization of the magnet is not used. The stability of the trapping potential is defined by the stability of the power supplies. The DC voltages for the electrodes of the measurement trap are applied from a single power supply module (ISEG, model EHS F210n) and demonstrate coherent changes over time.

\begin{figure*}[htb]
\centering
\includegraphics[width=0.95\textwidth]{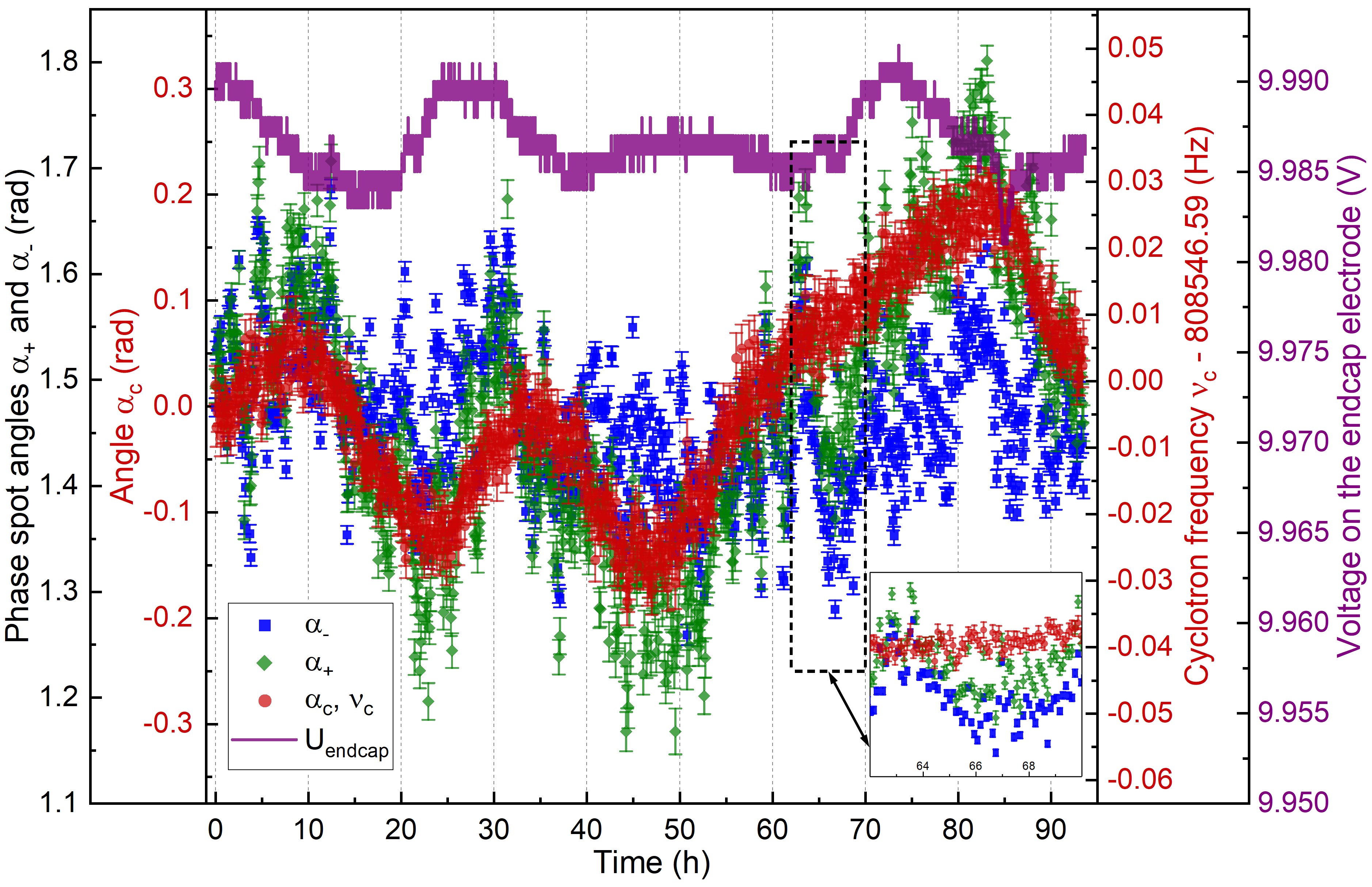}
\caption{The polar angles $\alpha_-$ and $\alpha_+$ of the magnetron (blue squares) and the cyclotron (green rhombuses) phase spots, respectively, on the detector for $^{133}$Cs$^+$ ions measured at JYFLTRAP with the PI-ICR method as a function of time. The corresponding angles $\alpha_c$ (left axis) and the cyclotron frequencies $\nu_c$ (right axis) are represented by red points. The phase accumulation time was 1 s. Readback value of the power supply that supplies the voltage to the endcap electrodes of the measurement trap is also shown (purple line). Inset on the bottom right shows the scatter of the individual phase spot angles $\alpha_\pm$ and their difference $\alpha_c$. It is evident that $\alpha_c$ is rather immune to scatter of $\alpha_\pm$ originating from voltage fluctuations.}
\label{fig:angles}
\end{figure*}

Figure~\ref{fig:angles} shows the polar angles and cyclotron frequencies of $^{133}$Cs$^+$ ions delivered from the surface ionization source, which were measured with the PI-ICR method for about 93.5 h at JYFLTRAP. For each single frequency measurement 700$-$800 ions were collected in total for both phase spots in about 7 minutes with the phase accumulation time $t_{acc}$ of 1~s. Up to 5 detected ions/bunch were taken into the analysis. The drift of the angle $\alpha_c$ and the cyclotron frequency $\nu_c$ (Eq.~\ref{eq:freq_alpha}) corresponds to the drift of the magnetic field $B$ and represents a smoother change in time compared to the polar angles $\alpha_-$ and $\alpha_+$, depending on the radial frequencies $\nu_-$ and  $\nu_+$, respectively, which are influenced in addition to the magnetic field by the trapping potential $U_0$. Daily fluctuations of the cyclotron frequency $\nu_c$ and the magnetron frequency $\nu_-$ do not exceed 30 mHz and 50 mHz, respectively.

\begin{figure}[htb]
\includegraphics[width=0.49\textwidth]{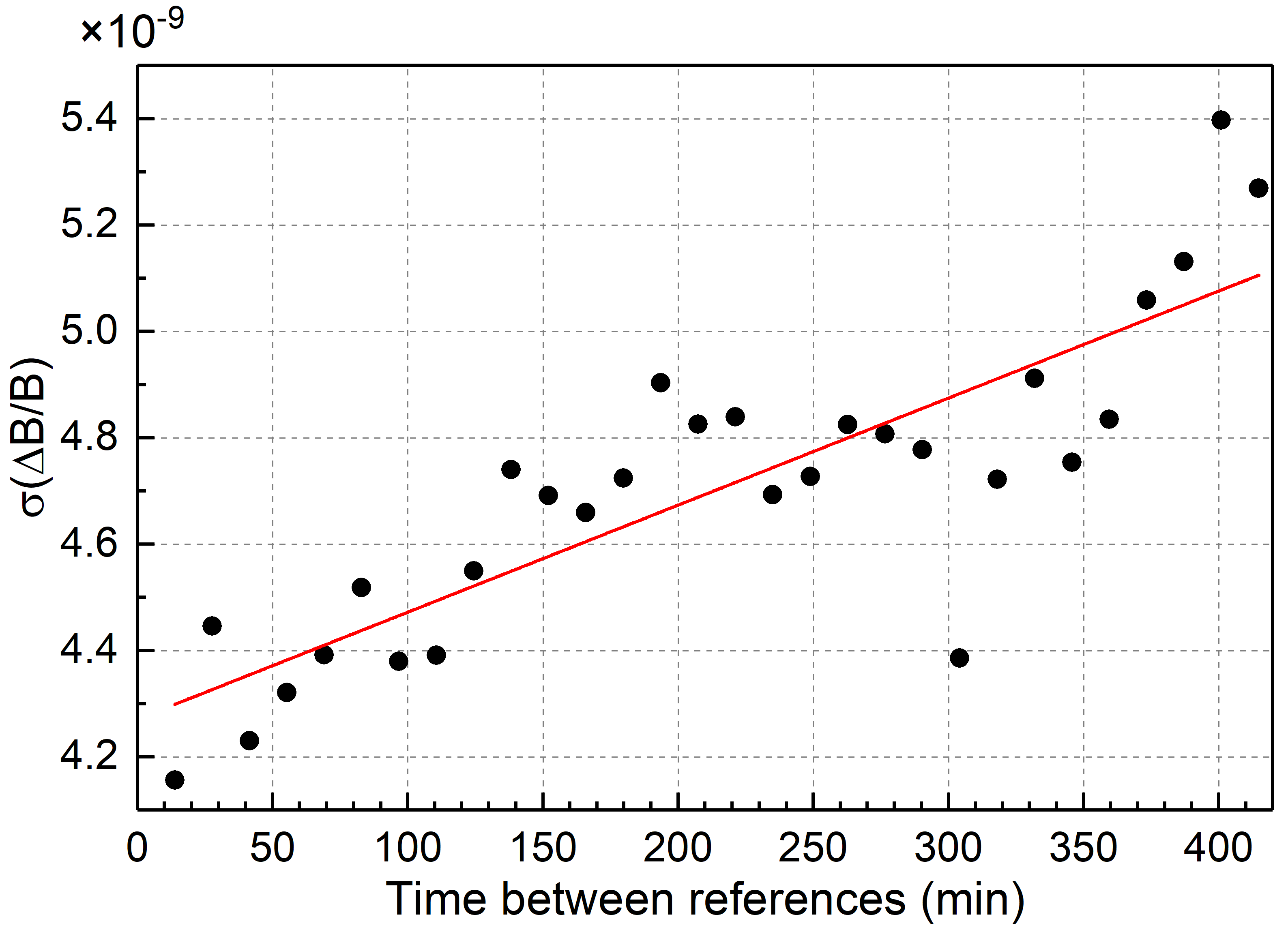}
\caption{Standard deviation of the relative magnetic-field fluctuations as a function of time between two reference measurements.}
\label{fig:B-error}
\end{figure}

To determine the uncertainty due to non-linear magnetic-field fluctuations the cyclotron frequency at time \\$t_{k+1}=(t_{k} + t_{k+2})/2$ was interpolated using Eq.~\ref{eq:interpolation} and compared to the actual value $\nu_c (t_{k+1})$. The relative deviation of the magnetic field from the linear change by using Eq.~\ref{eq:qbm} can be written as
\begin{multline} \label{eq:B-fluctuation}
\frac{\Delta B(t_{k+1})}{B(t_{k+1})} = \frac{B^{inter}(t_{k+1}) - B(t_{k+1})} {B(t_{k+1})} =\\ \frac{\nu_c^{inter}(t_{k+1}) - \nu_c(t_{k+1})} {\nu_c(t_{k+1})}.
\end{multline}
The standard deviation of $\Delta B(t_{k+1})/B(t_{k+1})$ was calculated for different time intervals $\Delta t = t_{k+2} - t_{k}$ and plotted as a function of $\Delta t$ (Fig.~\ref{fig:B-error}). The data were fitted by a straight line, the slope of which describes the uncertainty related to temporal fluctuations of the magnetic field per time interval between the reference measurements. It results in the standard uncertainty 
\begin{equation} \label{eq:B-error}
\frac{\delta B}{B \delta t} = 2.01(25) \times 10^{-12} \textnormal{ min}^{-1}.
\end{equation}
In previous studies of the magnetic field fluctuations at JYFLTRAP the cyclotron frequency measurements for $^{84}$Kr$^+$ ions were performed using the TOF-ICR method with Ramsey excitation pattern of 25-350-25 ms and the value\\ $\delta B/(B \delta t) = 8.18(19) \times 10^{-12}$ min$^{-1}$ was obtained \cite{CanetePhD}. We also note that temperature stabilization in the magnet bore and the pressure stabilization in the liquid-helium cryostat implemented at the SHIPTRAP setup (GSI), which has the superconducting magnet similar to the JYFLTRAP magnet, allowed to reach the value $\delta B/(B \delta t) = 1.3(11) \times 10^{-12}$ min$^{-1}$ \cite{Droese2011}.

\subsection{Contaminations and magnetron phase correction\label{sec:magnetron_phase}}

Ideally only one ion species should be present in the measurement trap during a cyclotron frequency measurement.
However, not always the mass-selective buffer gas cooling technique \cite{Savard1991} in the preparation trap or the Ramsey cleaning method \cite{Eronen2008} can provide a pure beam. For example, in the case of low-lying isomers the PI-ICR measurement is performed with more than one ion species in the trap \cite{Vilen2020b,Nesterenko2020b}.
The resolving power of the PI-ICR method is defined by the ability to resolve the phase difference of the cyclotron motions between two ion species accumulated during a phase accumulation time $t_{acc}$ \cite{Eliseev2014,Nesterenko2018}. It allows to separate the isomeric states of short-living isotopes ($T_{1/2} < 1$~s) with energy difference of a few tens of keV, which will have different angular positions of the cyclotron spots on the detector.

Due to finite excitation pulse duration of the dipolar $\nu_+$ and quadrupolar $\nu_c$ excitations (Fig.~\ref{fig:excitation_scheme}), the different ion species accumulate a small phase difference already before the actual phase accumulation period $t_{acc}$.
With only single ion species in the trap, this pre-accumulation has no effect as it merely adds a constant phase for both magnetron and cyclotron phase spots. However, when multiple ion species are present, this effect becomes significant for pattern 1 of Fig.~\ref{fig:excitation_scheme}. 
The position of the magnetron spot represents the average position of magnetron phases of all ion species present in the trap weighted by their fractions. Typically the differences of the phases are smaller than the width of the ion distribution and thus there is no way to distinguish the individual magnetron phase spots. Nevertheless, there is a way to correct for this effect.

Let us consider how the phase difference of two ion species $\Delta \phi = \phi_2 - \phi_1 $ changes during the complete PI-ICR measurement, marking with index ``1'' and ``2'' the parameters related to the first and the second ion species, respectively.
According to Eq.~(\ref{eq:dipolar_exc_phase}), the phase difference accumulated during the dipolar excitation time $t_+$ at the frequency $\nu_d$ (see Fig.~\ref{fig:excitation_scheme} step 3) is $\Delta \phi_+ = \pi (\nu_+^{(2)} - \nu_+^{(1)}) t_+$. To note, the accumulation of phase difference $\Delta \phi_+$ is only half of that of the free cyclotron motion.
After the excitation, during the delay time $t_{d1}$ (see Fig.~\ref{fig:excitation_scheme} between step 3 and step 4) the ions freely rotate at the modified cyclotron frequency and additionally accumulate the phase difference $\Delta \phi_{f} = 2 \pi (\nu_+^{(2)} - \nu_+^{(1)}) t_{d1}$. Then, the quadrupolar excitation pulse with duration of $t_c$ (see Fig.~\ref{fig:excitation_scheme} step 4) at the frequency $\nu_q$ is applied converting the cyclotron motion into the magnetron motion. Using Eq.~(\ref{eq:conversion_exc_phase}) and taking into account that $\nu_-^{(1)} = \nu_-^{(2)}$, the phase difference accumulated during conversion pulse is then given by $\Delta \phi_{c} = \pi (\nu_c^{(1)} - \nu_c^{(2)}) t_{c}$. Note that the conversion pulse also inverts the phase of the cyclotron motion accumulated earlier. The total phase difference obtained by combining the phase differences from each of the three stages can be written as
\begin{equation} \label{eq:mag_phase_diff}
\Delta \phi = 2 \pi \left( \nu_c^{(1)} - \nu_c^{(2)} \right ) \left( \frac{t_+}{2} + t_{d1} + \frac{t_c}{2} \right),
\end{equation}
assuming $\nu_+^{(1)} - \nu_+^{(2)} = \nu_c^{(1)} - \nu_c^{(2)}$. As can be seen from this expression, the cyclotron phase difference is accumulated twice as slower during excitations compared to the free cyclotron rotation.
Similarly to the correction in \cite{Orford2020}, the correction angle for the ion species of interest $i$ with the cyclotron frequency $\nu_c^{i}$ added to the measured angle of the magnetron phase spot $\alpha_-$ can be written as
\begin{equation} \label{eq:mag_phase_diff}
\alpha_{corr}^i = 2 \pi t_{corr} \sum_{k=1}^N \chi_k \left( \nu_c^{(i)} - \nu_c^{(k)} \right),
\end{equation}
where $t_{corr} = (t_+ / 2 + t_{d1} + t_c / 2$), and $\chi_k$ and $\nu_c^{k}$ are the relative population fraction and cyclotron frequency of the $k^{th}$ ion species, respectively. The fractions of the different ion species can be deduced from the separated cyclotron phase spots of different ion species on the detector. The cyclotron frequencies of the ion species can be determined from preliminary analysis before applying the correction for the magnetron phase.

\begin{figure}[t]
\includegraphics[width=0.49\textwidth]{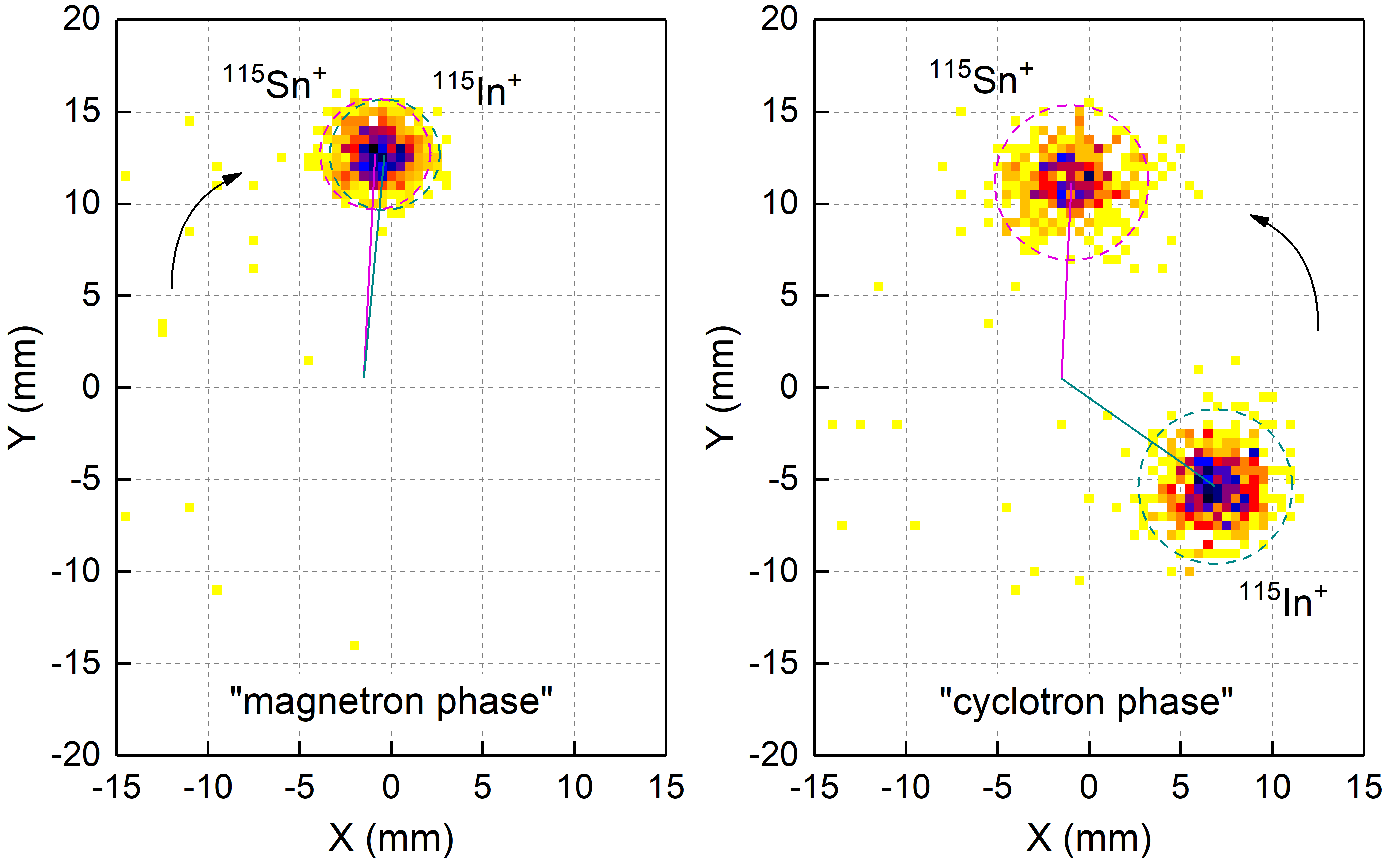}
\caption{Projection of the phase spots for $^{115}$Sn$^+$ and $^{115}$In$^+$ ions, simultaneously stored in the measurement trap, on the position-sensitive detector for a single cyclotron frequency measurement in the PI-ICR method with the phase accumulation time of 307 ms. While the cyclotron phase spots are clearly separated, the magnetron phases are indistinguishable on the detector, but yet they have a significant angle requiring a correction.}
\label{fig:Sn-In-PIICR}
\end{figure}

In the discussion above, it was assumed that the magnetron frequency is mass independent. However, it has a very weak dependency on mass as can be seen starting with the second-order series expansion in $\nu_z^2 / \nu_c^2$ of Eq.~(\ref{eq:radial_freq}). In the series expansion up to the third order, the magnetron frequency $\nu_-$ is approximated as
\begin{equation} \label{eq:mag_freq}
\nu_- \approx \frac{U_0}{4 \pi B d^2} + \frac{U_0^2}{8 \pi B^3 d^4} \frac{m}{q} + \frac{U_0^3}{8 \pi B^5 d^6} \left( \frac{m}{q} \right)^2.
\end{equation}
Using the JYFLTRAP parameters ($U_0$ = 100~V, $B$ = 7~T, $d^2 = 6.9 \times 10^{-4}$ m$^2$), the magnetron-frequency difference $\nu_-^{(2)} - \nu_-^{(1)}$ between two singly-charged ions with the mass difference $\Delta m = m_2 - m_1$ can be written as
\begin{multline} \label{eq:mag_freq_diff}
\frac{\nu_-^{(2)} - \nu_-^{(1)}}{[Hz]} \approx \frac{U_0^2}{8 \pi B^3 d^4} \frac{\Delta m}{q} + \frac{U_0^3}{8 \pi B^5 d^6} \frac{m_2^2 - m_1^2}{q^2} \approx \\
3 \cdot 10^{-2} \cdot \frac{\Delta m}{[u]} + 8 \cdot 10^{-7} \cdot \frac{(m_1 + m_2) \Delta m}{[u]^2}.
\end{multline}
The first term is dominant in Eq.~(\ref{eq:mag_freq_diff}) for all ions used at JYFLTRAP. For example, the mass difference of 1 MeV/$c^2$ corresponds to the magnetron-frequency difference of about $3 \times 10^{-5}$ Hz. This frequency shift results in the error of the cyclotron frequency ratio $R$ at the level well below $10^{-10}$ for typically used ions and can be neglected. Ions with larger than 1 MeV/$c^2$ mass differences can be separated relatively easily using the Ramsey cleaning method \cite{Eronen2008} or the mass-selective buffer gas cooling technique \cite{Savard1991}, which allows to perform the measurements with a single ion species, thus, eliminating the shift completely. For smaller mass differences the effect is so small that it is safe to assume that the magnetron frequency is mass independent with a sufficient accuracy for all practical cases at JYFLTRAP.

So far, typical pulse durations of the dipolar and quadrupolar excitations used at JYFLTRAP have been 1~ms and 2~ms, respectively. The usual delay time between excitation pulses $t_{d1}$ is about 50~$\mu$s. However, there is a room to reduce all of them. Figure~\ref{fig:Sn-In-PIICR} shows the projections of the magnetron and cyclotron phases of $^{115}$Sn$^+$ and $^{115}$In$^+$ ions simultaneously stored in the measurement trap. The atomic mass difference between $^{115}$In and $^{115}$Sn was measured with the FSU cryogenic Penning trap mass spectrometer with a high accuracy ($\Delta m = 497.489(10)$ keV/$c^2$ \cite{Mount2009}), giving the cyclotron frequency difference $\Delta \nu_c = \nu_c(^{115}\textnormal{Sn}^+) - \nu_c(^{115}\textnormal{In}^+) =$ 4.34697(9)~Hz for singly-charged ions at JYFLTRAP. The phase accumulation time used in the measurement was about 307~ms and the accumulated phase difference for the cyclotron phases $^{115}$Sn$^+$ and $^{115}$In$^+$ was one full turn plus about 120$^{\circ}$. According to Eq.~\ref{eq:mag_phase_diff} the phase difference of the two ion species in the magnetron spots is about of 2.4$^{\circ}$. Taking into account the correction for the magnetron phase spot and using the count-rate class analysis (see Sec.~\ref{sec:countrate}) to account ion-ion interactions in the trap, the result of the measurement with 18 cyclotron-frequency ratios is in an agreement with the FSU $Q$-value \cite{Mount2009}, giving $\Delta m = 497.43(57)$ keV/$c^2$ for the mass difference of $^{115}$In and $^{115}$Sn. Without the correction of the magnetron spot position, the mass difference $\Delta m = 499.64(57)$ keV/$c^2$, differs from FSU \cite{Mount2009} by more than 2~keV/$c^2$, i.e., by more than 3~$\sigma$. Note, that the cyclotron frequency measurements were performed alternately with the settings for $^{115}$Sn and $^{115}$In in such a way that the angle between the magnetron and cyclotron phases of the ion of interest was close to zero in order to reduce the error due to distortion of the projection and conversion error.

To reduce the effect of the overlapping magnetron spots, the durations of the excitation pulses should be minimized. The main limitation concerns the conversion pulse, which typically has an amplitude of 5.6 V for 2 ms of duration at JYFLTRAP, and can be shortened to about of 1.12 ms (function generators provide 10 V for the maximum amplitudes).
Another approach is to increase the delay time $t_{d1}$ in such a way that the phases obtained in pattern 1 of excitation scheme are also separated on the detector. However, in this case, the phase accumulation time $t_{acc} = t_{d2} - t_{d1}$ is reduced for the same trap cycle length, reducing the accuracy of the cyclotron frequency determination.

\subsection{Ion-number dependence}

More than one ion species in the trap during a TOF-ICR or PI-ICR measurement introduces also frequency shifts. If ions of a single species are simultaneously stored in the trap the space charge of such a cloud of ions does not influence the center-of-mass motion of the cloud. A driving field acts on the mass center of the cloud and no frequency shifts are observed \cite{Bollen1990,Bollen1992}. However, if too many ions are stored in the trap and the space charge pushes a significant portion of ions from the center region of the trap to the region of the field imperfections, it leads to a frequency shift. 
In case of more than one ion species confined in the trap, the center-of-mass motions of ion species can interact with each other causing large shifts in the cyclotron frequencies. The frequency shift depends on the total number of stored  ions, the difference of the cyclotron frequencies and the ratio between the number of the ions of interest and contaminant ions \cite{Bollen1992}.

\subsubsection{Count-rate class analysis\label{sec:countrate}}

In order to identify and correct for the cyclotron frequency shift due to the presence of contaminant ions in the measurement trap, the count-rate class (z-class) analysis can be performed \cite{Kellerbauer2003}. The frequency data are divided into classes by the count rate, i.e. the number of ions simultaneously stored in the trap. The frequencies determined for each class are plotted as a function of the center of gravity of the ions in the class and fitted by a straight line. Extrapolation to the efficiency of the detector results in the corrected frequency. A similar analysis can be directly performed with the cyclotron frequency ratios \cite{Roux2013}.

\begin{figure}[htb]
\includegraphics[width=0.49\textwidth]{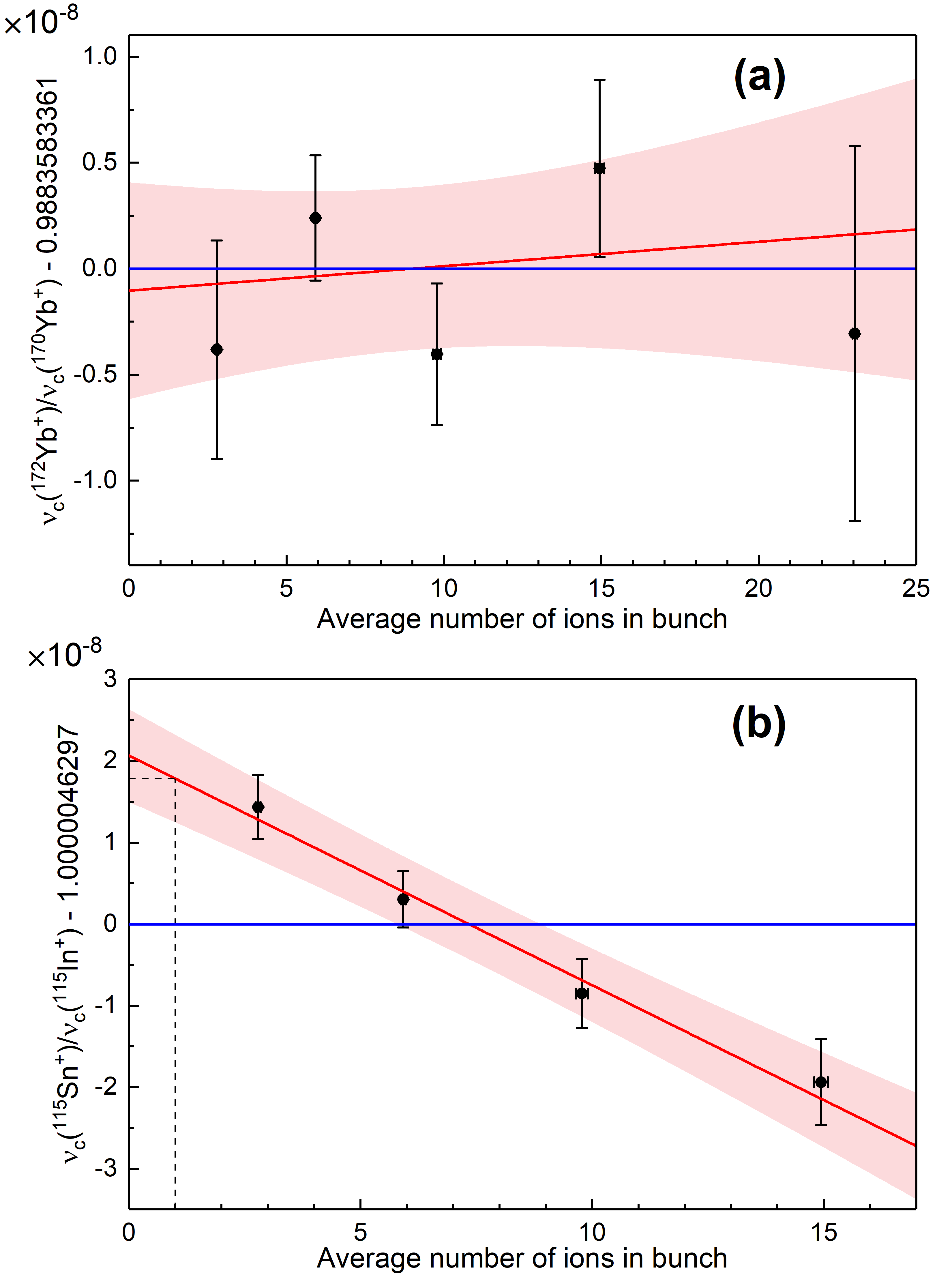}
\caption{Cyclotron frequency ratio as a function of the number of ions in the measurement trap. The detected number of ions per bunch was corrected according to the efficiency of the detector. The red line and the red shaded band represent a linear fit and 1$\sigma$ uncertainty of the fit, respectively. The blue line is the mean frequency value, averaged over all ions per bunch. (a) The cyclotron frequency ratios of $^{172}$Yb$^+$ and $^{170}$Yb$^+$ ions are shown for the measurement with only one ion species simultaneously stored in the trap. (b) The cyclotron frequency ratios of $^{115}$Sn$^+$ and $^{115}$In$^+$ ions are shown for the measurement when both ion species were simultaneously in the trap.}
\label{fig:z-class}
\end{figure}

In our case, as shown below (Sec.~\ref{sec:efficiency}), the detector efficiency $\epsilon$ depends on the count rate, it is not a constant. Therefore, the frequencies or frequency ratios are plotted as a function of number of detected ion per bunch corrected by the detector efficiency and, then, extrapolated to the unity, i.e. a single ion in the measurement trap \cite{Vilen2019}. Figure~\ref{fig:z-class}a shows a dependence of the cyclotron frequency ratio $\nu_c$($^{172}$Yb$^+$)/$\nu_c$($^{170}$Yb$^+$) measured with the PI-ICR method on the number of ions per bunch in the trap. The count-rate classes with 1 to 5 detected ions per bunch, corrected by the efficiency $\epsilon$, were used. Only the ions of single species, either $^{172}$Yb$^+$ or $^{170}$Yb$^+$, were present in the measurement trap at the same time and no frequency shift was observed. A linear fit (red line) is in agreement with the mean frequency value (blue line), averaged over all classes.

The situation when two ion species are simultaneously stored in the trap is demonstrated for $^{115}$Sn$^+$ and $^{115}$In$^+$ ions, created at the same time from tin-indium alloy electrode in the glow discharge ion source. Two ion species were resolved in the PI-ICR measurement, which corresponds to two different cyclotron phase spots on the detector (Fig.~\ref{fig:Sn-In-PIICR}). 
The data were divided into four groups according to the number of detected ions per bunch and corrected by the efficiency $\epsilon$. The cyclotron frequency of the of $^{115}$Sn$^+$ ions decreases with increasing the number of ions in the trap, while the cyclotron frequency of the $^{115}$In$^+$, on the contrary, increases. Thus, the frequencies of the states shifted towards each other and the magnitude of the shift depends on the number of stored ions. Figure~\ref{fig:z-class}b shows this dependence for the cyclotron frequency ratio $\nu_c$($^{115}$Sn$^+$)/$\nu_c$($^{115}$In$^+$). The frequency ratio was extrapolated to a single ion in the trap by using a linear fit, represented by the red line with a red band of 1$\sigma$ uncertainty of the fit. Note that the data points have also $x-$axis error bars due to the uncertainty of the efficiency curve of the detector and an orthogonal distance regression fitting method was used.

The analysis performed for the cyclotron-frequency ratio measurement of $^{115}$Sn$^+$ and $^{115}$In$^+$ ions (see Sec.~\ref{sec:magnetron_phase}) with the data limited to 1-4 detected ions/bunch and without count-rate class analysis results in the mass difference $\Delta m = 495.25(29)$ keV/$c^2$ of $^{115}$In and $^{115}$Sn, which differs from FSU $Q$-value \cite{Mount2009} by $-2.24(29)$ keV. The difference between the cyclotron frequency ratios received without and with count-rate class analysis is $-2.04(60)\times 10 ^ {-8}$.

\subsubsection{Efficiency of the MCP detector\label{sec:efficiency}}

The position-sensitive ion detector used in the PI-ICR measurements at JYFLTRAP is a microchannel plate (MCP) detector with two plates in chevron configuration with a delay-line anode (RoentDek GmbH, model DLD40). Data acquisition system (RoentDek GmbH, model TDC8HP) operates with software developed locally at JYFLTRAP, using the drivers provided by the manufacturer. Ion hit events consist of a MCP backplate signal and of four delay line signals, which are needed to reconstruct the position. Currently the data acquisition software is realized so that even if just one of the delay line signals is missing, this event is discarded in the PI-ICR analysis.
The MCP detector is located outside the superconducting magnet at a distance of 104 cm from the center of the measurement trap, where the magnetic field strength $B$ is about of 30 mT.

\begin{figure}[htb]
\includegraphics[width=0.49\textwidth]{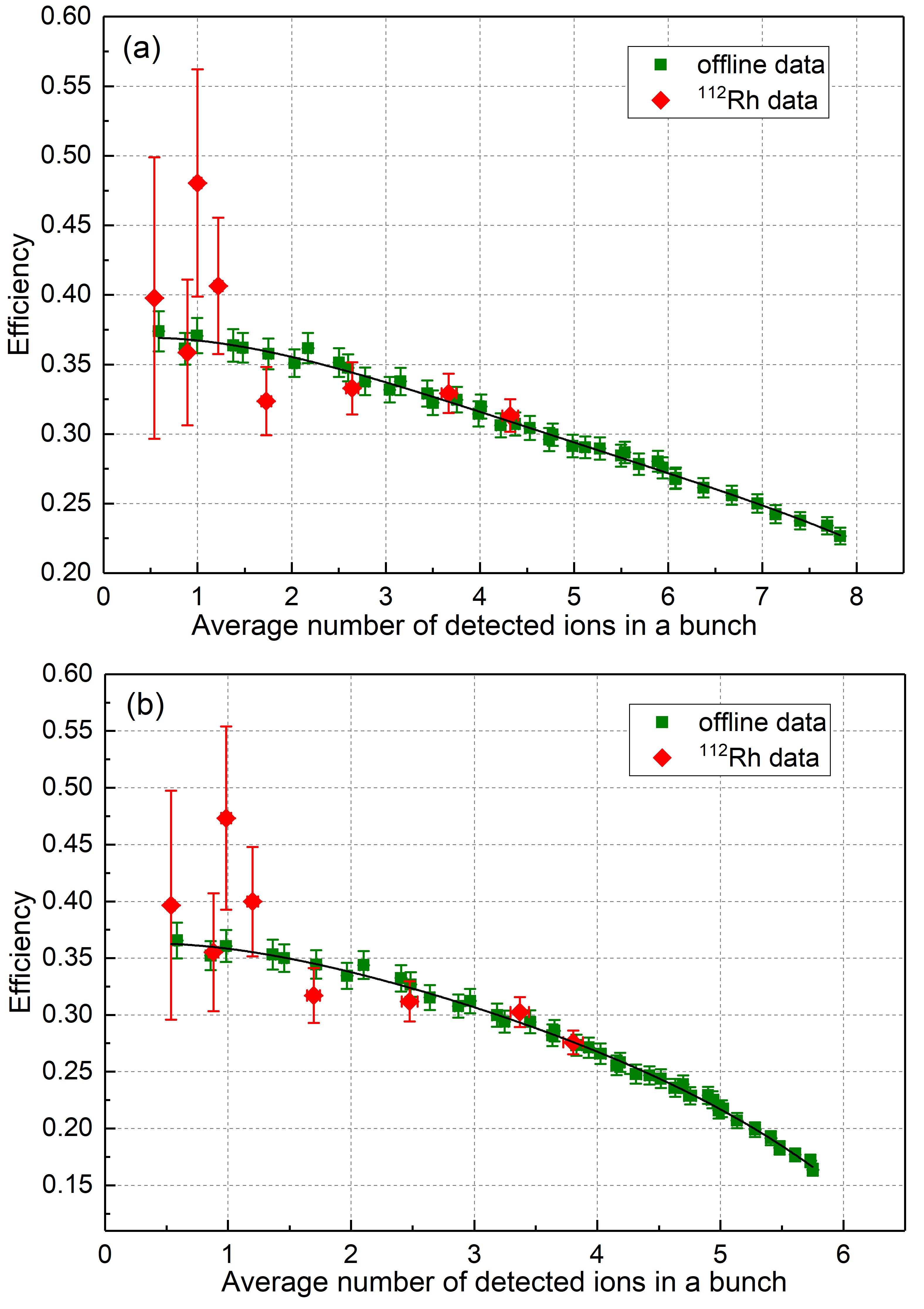}
\caption{Efficiency of the MCP detector as a function of the detected number of ions in a bunch. Red points indicate the online data obtained by comparing a count rate of $^{112}$Rh$^+$ ions on the MCP detector and a beta-decay rate on the silicon detector with known efficiency. Green points indicate the offline data obtained with $^{133}$Cs$^+$ ions entering the RFQ by scanning its opening time. (a) Efficiency for the data with time-of-flight information. (b) Efficiency for the data with time-of-flight and successful ion position reconstruction.}
\label{fig:MCPefficiency}
\end{figure}

The measurement of the MCP detector efficiency was based on a comparison of the count rate of radioactive ions on the MCP detector and the silicon detector (300 mm$^2$ of area and 500 $\mu$m of thickness) with a known efficiency of 30.72(28)\% \cite{Khanam2017}. The MCP is placed on an actuator and can be removed from the beam axis to pass the beam downstream to the silicon detector located 41 cm behind the MCP on the same beam axis. Thus, the measurement can be performed alternately with both detectors. 
$\beta$-decaying ions of $^{112}$Rh$^+$ were used to measure the MCP efficiency. They were produced in proton-induced fission of $^{nat}$U at IGISOL, purified from the isobars in the preparation trap of JYFLTRAP and sent to either the MCP or silicon detector. Mainly isomeric state of $^{112}$Rh with T$_{1/2}$ = 6.73 (15) s is populated in fission reactions. Data were collected for 180 s alternately on both detectors for different intensities of the radioactive beam. 
The daughter of $^{112}$Rh is $^{112}$Pd having $T_{1/2} \approx 21$ h and effectively does not contribute to $\beta$-background. The $^{112}$Rh activity was let to saturate and this activity was taken as the rate of ions after correcting for $\beta$-efficiency.
Red points in Fig.~\ref{fig:MCPefficiency} represent the efficiency based on the number of detected ions on the MCP detector and beta-decays on the silicon detector as a function of the average number of detected ions in the ion bunches.

Additionally to the online data with $^{112}$Rh$^+$ ions, the offline measurements with the stable $^{133}$Cs$^+$ ions, produced at the offline source station (Fig.~\ref{fig:IGISOL}) before the dipole magnet, were performed to get better statistics and extend the efficiency curve. 
Pulsing the voltage on a kicker electrode, located at the electrostatic switchyard, allows to control the time (beam gate) when the ions let to enter the RFQ. Thus, the number of ions in bunches formed in the RFQ is proportional to the beam gate time. This time was scanned and the number of ions, detected on the MCP after cooling and centering in the preparation trap, provided information about the relative MCP efficiency. The data obtained in this way (green points in Fig.~\ref{fig:MCPefficiency}) were fitted by a polynomial of 4th order and scaled with a constant factor to match the absolute efficiency (red points). The efficiency curve shows a significant decrease with an increase in the number of ions in the bunch. 
Figure~\ref{fig:MCPefficiency}a shows the data with time-of-flight information (signal from the MCP backplate) for the ions, while Figure~\ref{fig:MCPefficiency}b shows the data with both time-of-flight information and the necessary information (all the four delay line signals) from the delay-lines to successfully reconstruct the ion impact position. With an increase in the number of ions per bunch, a loss of signal from one or several outputs of the delay-line anode is observed, which leads to a failed reconstruction of the ion impact position on the detector. Likely this due to pileup of signals and dead time of the TDC8HP.

The measured MCP efficiency (Figure~\ref{fig:MCPefficiency}) is higher than the efficiency determined earlier, when the MCP was located closer to the magnet \cite{Vilen2019}. In addition to reducing the effect of the magnetic field, this can also be significantly due to an increase in the time width of the ion bunches arriving at the MCP. The voltages on the extraction electrodes were changed to obtain the same magnification factor \cite{Nesterenko2018} as before the movement of the MCP, which in turn led to an increase in the width of the time-of-flight distribution of ions on the detector. With wider time-of-flight distribution, the effects of pile-up and dead time of the TDC are smaller.

\subsection{Cross-reference measurements with carbon clusters\label{sec:clusters}}

\begin{figure}[t]
\includegraphics[width=0.49\textwidth]{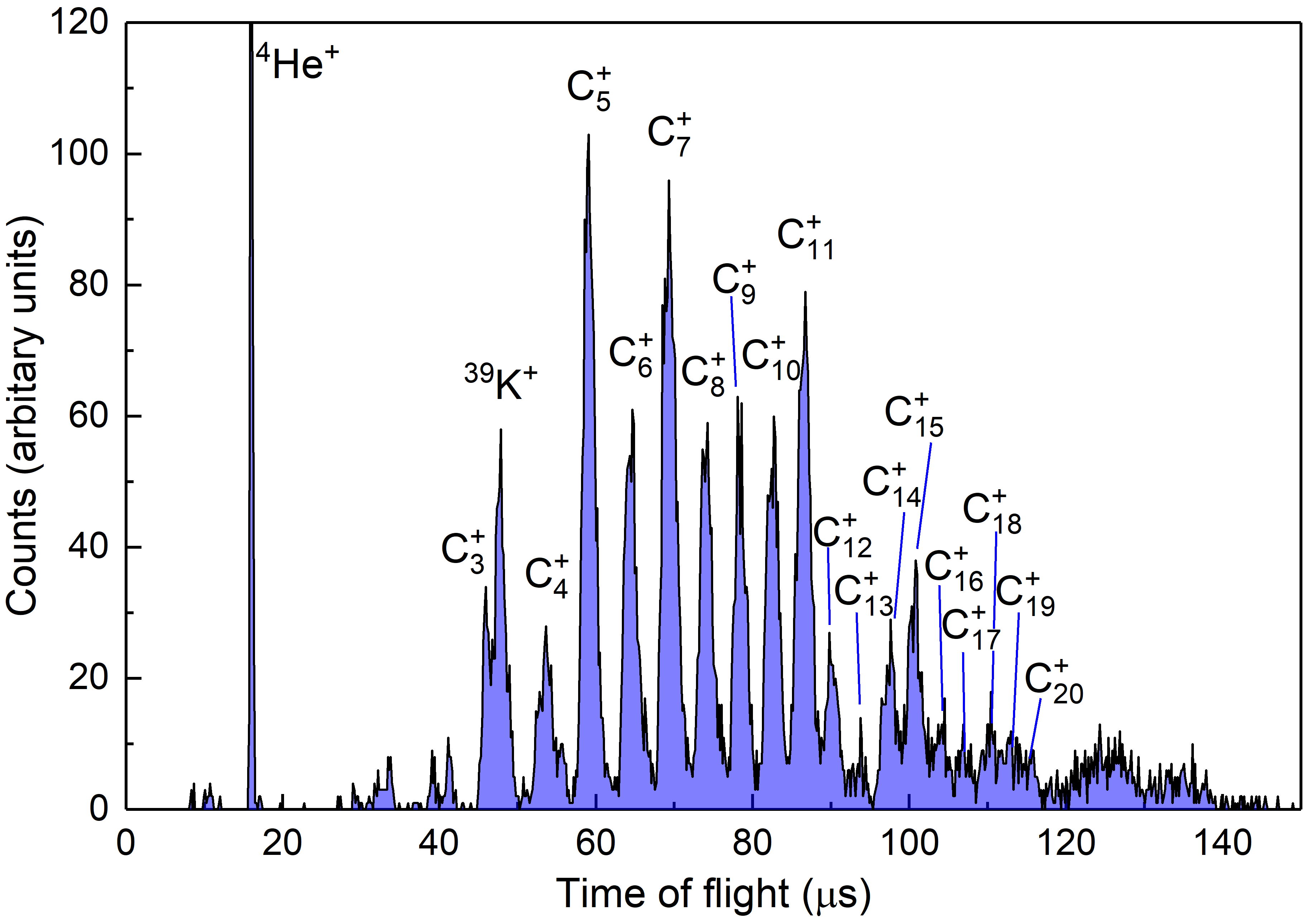}
\caption{Time-of-flight spectrum of carbon cluster ions detected at the MCP after trapping and cooling for 265 ms in the preparation trap. No rf excitations applied in this example. The open time of the preparation trap was scanned from 100~$\mu$s to 500~$\mu$s.}
\label{fig:clusters}
\end{figure}

In this study a Sigradur\textsuperscript{\textregistered} glassy carbon plate (disc-shaped, 16~mm diameter, 2~mm thickness) was used to produce carbon clusters in the laser ablation ion source (Sec.~\ref{sec:ion_source}). Laser pulse creating a bunch of ionized clusters was synchronized with the open time of the preparation trap to capture the ions. Figure~\ref{fig:clusters} shows a time-of-flight spectrum of singly-charged carbon clusters detected at the MCP. The open time of the preparation trap was scanned from 100~$\mu$s to 500~$\mu$s. The ions were trapped and cooled for 265~ms in the preparation trap without any rf excitations and then extracted through the measurement trap to the MCP. The spectrum shows mainly carbon clusters $^{12}$C$_n$ with $n$ from 3 up to at least 20. The main contaminants are He$^+$ and K$^+$ ions.

\begin{figure}[t]
\includegraphics[width=0.49\textwidth]{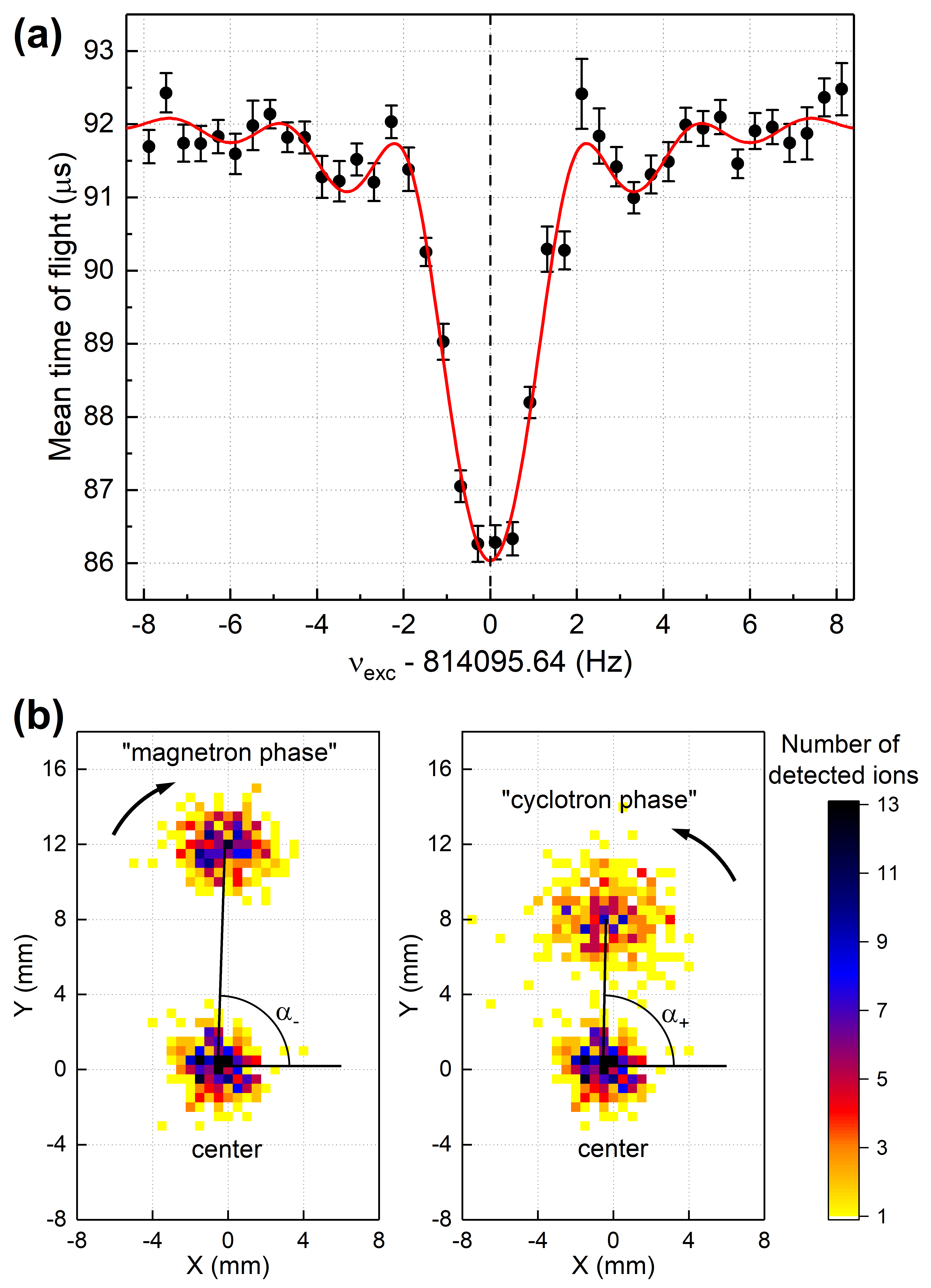}
\caption{Single cyclotron frequency measurements performed for $^{12}$C$^+ _{11}$ carbon cluster ions at JYFLTRAP. (a) TOF-ICR spectrum obtained with 400 ms of excitation time. The black points with error bars, represented the mean time-of-flight for each scanned frequency, are fitted with the theoretical curve \cite{Konig1995} (red line). (b) Projection of the trap center and accumulated phase spots on the position-sensitive detector for a single cyclotron frequency measurement in the PI-ICR method with the phase accumulation time 400 ms.}
\label{fig:C11-measurement}
\end{figure}

In the measurement cycle the bunches of the produced carbon cluster ions were captured into the preparation trap. The optimum capture time was chosen to select the desired cluster size. The ions were cooled and purified in the preparation trap for about 235 ms and transferred into the measurement trap. Then the purified ions of interest were transported back to the preparation trap for an additional step of cooling for 133 ms. 
One step of cooling was insufficient for effective cooling and centering of the ions due to large number of ions of different cluster sizes.
Finally, the ions of interest were sent to the measurement trap, where the cyclotron frequency determination using the PI-ICR method with the phase accumulation time of 400 ms was performed (see Fig.~\ref{fig:C11-measurement}b). Both traps worked in parallel in such a way that when the cyclotron frequency of ions was determined in the measurement trap, a new ion bunch was prepared in the preparation trap. A single cyclotron frequency measurement took about 10 minutes.

Ion interaction with the residual gas in the measurement trap results in a damping of the radial motions and increase in the size of the images of the phase spots. The main source of the residual gas is the flow of helium atoms entering from the preparation trap through a pumping barrier into the measurement trap. Since the modified cyclotron frequency $\nu_+$ is much higher than the magnetron frequency $\nu_-$, the ions performing the cyclotron motion travel a longer distance and undergo more collisions with the residual gas atoms than the ions performing the magnetron motion during the phase accumulation time $t_{acc}$. Thus, the damping effect is manifested stronger for the cyclotron phase spot with an increase in $t_{acc}$ than for the magnetron phase spot (Fig.~\ref{fig:C11-measurement}b). A collision of an ion, performing a radial motion, with the residual gas results in a decrease in the cyclotron motion radius and an increase in the magnetron motion radius \cite{Eliseev2014}. The radial frequencies are slightly shifted due to damping, however, their sum is still the cyclotron frequency $\nu_c$ \cite{Kretzschmar2008}. The effects of damping and smearing of the cyclotron motion increase the statistical uncertainty of the angle $\alpha_+$.
The damping effect for the carbon clusters is stronger than for monatomic ions. Damping effects for the ion motions in a Penning trap can be described by introducing a damping force $\vec{F} = -2m \gamma \vec{\upsilon}$, where $\vec{\upsilon}$ is the ion velocity and $\gamma$ is the damping coefficient \cite{Kretzschmar2008,George2011}, 
\begin{equation} \label{eq:damping}
\gamma = \frac{q}{2m} \frac{1}{K_0} \frac{(p/p_0)}{(T/T_0)},
\end{equation}
where the reduced ion mobility $K_0$ is tabulated at normal atmospheric pressure $p_0 = 10^5$~Pa and room temperature $T_0 = 300$~K. The reduced ion mobility for the $^{133}$Cs$^+$ and $^{85}$Rb$^+$ ions in helium is about 18 and 20~cm$^2$s$^{-1}$V$^{-1}$ \cite{Viehland2012}, respectively, and it ranges from about 5~cm$^2$s$^{-1}$V$^{-1}$ for $^{12}$C$^+ _{20}$ ions to 12~cm$^2$s$^{-1}$V$^{-1}$ for $^{12}$C$^+ _{6}$ ions \cite{Helden1993,Shvartsburg1998}. Hence, the $\gamma$ is larger for the carbon cluster ions and they are more affected by residual gas than the typically used monoatomic ions. For example, the ratio of the cyclotron-to-magnetron motion radii $r_+/r_-$ in the PI-ICR measurement with 400 ms of the phase accumulation time was smaller by 22~\% for the $^{12}$C$^+ _{11}$ ions (A=132) compared to the $^{133}$Cs$^+$ ions with the similar mass number at the same helium gas flow in the preparation trap.

\begin{table*}[htb]
\caption{Carbon-cluster cross-reference measurements performed at JYFLTRAP with the PI-ICR method. The ions $^{12}$C$^+ _{9}$, $^{12}$C$^+ _{11}$ and $^{12}$C$^+ _{13}$ were chosen as the reference ions. The number of individual cyclotron frequency ratios $r_i = \nu_c(^{12}$C$^+ _{n, ref}) / \nu_c(^{12}$C$^+ _{n, ioi})/$ measured for the ions of interest is shown for each pair.}
\label{tab:clusers}
\begin{center}
\begin{tabular}{|c|c?c|c|c|c|c|c|c|c|c|c|}
\hline
 & $^{12}$C$^+ _{n, ioi}$ & 6 & 7 & 8 & 9 & 10 & 11 & 12 & 13 & 14 & 15 \\
\hline
$^{12}$C$^+ _{n, ref}$ & A & 72 & 84 & 96 & 108 & 120 & 132 & 144 & 156 & 168 & 180 \\
\Xhline{2\arrayrulewidth}
9 & 108 & 36 & 27 & 26 &  & 27 &  & 49 &  &  &  \\
\hline
11 & 132 &  &  & 26 & 45 & 22 &  & 23 & 29 & 28 &  \\
\hline
13 & 156 &  &  &  &  & 32 &  & 29 &  & 44 & 23 \\
\hline
\end{tabular}
\end{center}
\end{table*}

The cyclotron frequency ratio measurements have been performed for the singly-charged carbon cluster ions $^{12}$C$^+ _{n}$ with $6 \leq n \leq  15$ in three series, where the ions $^{12}$C$^+ _{9}$, $^{12}$C$^+ _{11}$ and $^{12}$C$^+ _{13}$ were used as reference ions (Table~\ref{tab:clusers}). A typical obtained statistical uncertainty for the frequency ratios was a few $\times 10^{-9}$. The uncertainty related to the magnetic field fluctuations (Sec. \ref{sec:magnet_stability}) was added quadratically to the statistical uncertainty and made a very minor contribution.
Count-rate class analysis for the measured cyclotron frequency ratios was performed \cite{Roux2013}. No dependence of the frequency ratio on the number of detected ions was observed and data with detected 1-5 ions/bunch were taken into account in the analysis.

The setup was optimized for a certain mass in the frequency ratio measurement. Especially the pressure of the purification trap is optimal for only a small mass range. Thus the masses significantly higher or lighter were not prepared in the preparation trap in an optimal way. Also, the damping effect was stronger for lighter carbon clusters at the same pressure in the trap, since the damping coefficient (Eq.~\ref{eq:damping}) slightly increases with decreasing the cluster size.

\begin{figure}[t]
\includegraphics[width=0.49\textwidth]{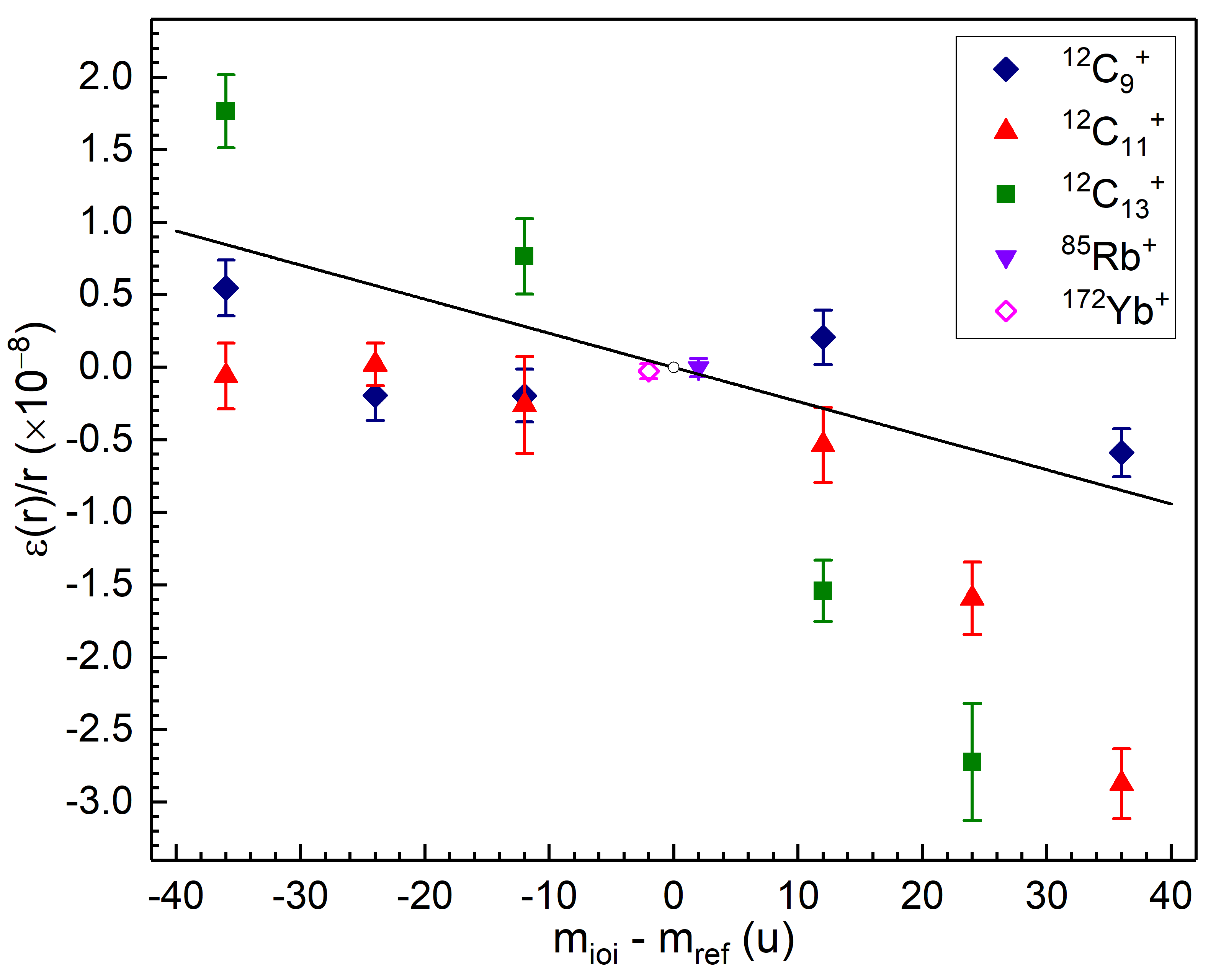}
\caption{Relative deviation of the measured cyclotron frequency ratios from the calculated ones as a function of the mass difference between reference ion and ion of interest. The straight line is a linear least-squares fit to the data. The measurements were performed with difference reference ions: $^{12}$C$^+ _{9}$ (blue rhombuses), $^{12}$C$^+ _{11}$ (red triangles), $^{12}$C$^+ _{13}$ (green squares), $^{85}$Rb$^+$ (violet circles) and  $^{172}$Yb$^+$ (pink circles).}
\label{fig:clusters-ratios}
\end{figure}

\begin{figure}[htb]
\includegraphics[width=0.49\textwidth]{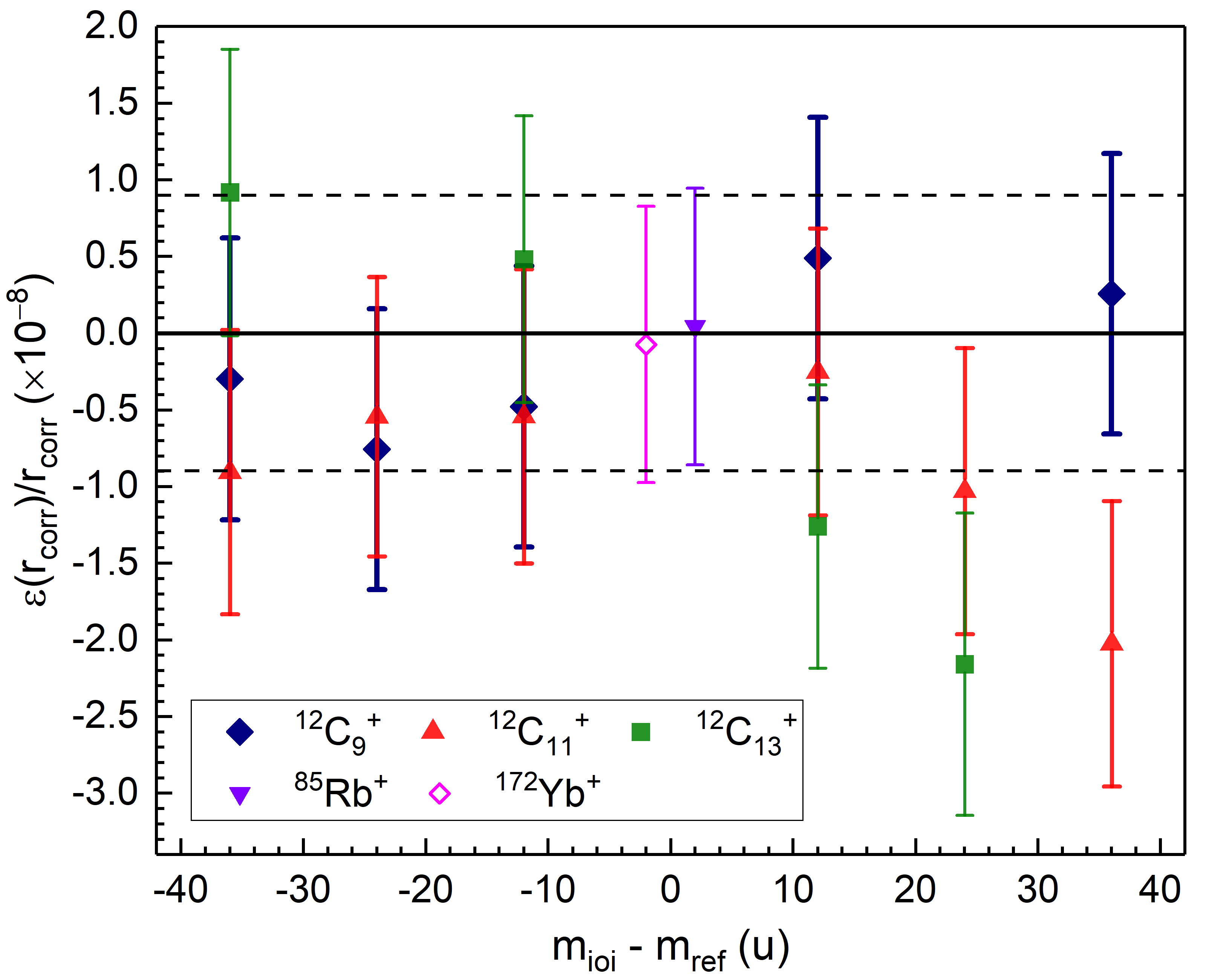}
\caption{Relative deviation of the measured cyclotron frequency ratios after correction for the mass-dependent shift. The dashed lines indicate the residual uncertainty which was added quadratically to the uncertainties of the cyclotron frequency ratios to obtain a reduced chi-square $\chi ^2 / N = 1$.}
\label{fig:clusters-ratios-corr}
\end{figure}

In addition to the measurements with the carbon clusters two previous PI-ICR measurements with $^{85,87}$Rb$^+$ and $^{170,172}$Yb$^+$ ions, which have very well-known mass values \cite{Wang2021}, were included in the analysis. The cyclotron frequency ratios $r=\nu_c(^{85}$\textnormal{Rb}$^+) / \nu_c(^{87}$\textnormal{Rb}$^+)$ \cite{Nesterenko2018} and\\ $r=\nu_c(^{172}$\textnormal{Yb}$^+) / \nu_c(^{170}$\textnormal{Yb}$^+)$ \cite{Nesterenko2020} were measured with a relative uncertainty of 0.64 and 0.5 ppb, respectively. 

The molecular binding energy (ionization energy) of the carbon clusters gradually changes in range from about of 5.3~eV to 6.6~eV per atom for the clusters C$_n$ with $6 \leq n \leq  15$ \cite{Tomanek1991,Manby1998}. Since the molecular binding energy is almost a constant, its contribution to the calculated cyclotron frequency ratio of the cluster ions is negligible, less than $10^{-10}$ in our cases. The binding energy of valence electron is about $9-10$~eV in the studied carbon clusters \cite{Belau2007}, about 4~eV in Rb$^+$ ions \cite{Lotz1970} and 6 eV in Yb$^+$ ions \cite{Lotz1970} and, thus, its contribution to the frequency ratio is also negligible. The mass of the singly-charged carbon cluster ions was calculated as $m(^{12}\textnormal{C}^+_n) = n \times m(^{12}\textnormal{C}) - m_e$, where an atomic mass of carbon $m$($^{12}$C)$=12$ u.

The weighted mean ratios $r$ of the measured individual cyclotron frequency ratios were compared with the calculated frequency ratios $r_{calc}$.
If the measurements of the cyclotron frequencies results in values which deviate from the correct frequencies by a constant offset, the relative shift of the cyclotron frequency ratio
\begin{equation} \label{eq:mass-shift}
\frac{\varepsilon(r)}{r} = \frac{r - r_{calc}}{r} \propto (m_{ref} - m_{ioi})
\end{equation}
is proportional to the mass difference of the reference ion $m_{ref}$ and ion of interest $m_{ioi}$. The cyclotron frequency offset, leading to a mass-dependent ratio shift, can be due to imperfections of the electric-quadrupolar field in a Penning trap or a misalignment of the electrostatic trapping field with respect to the magnetic field axis \cite{Bollen1996}.

The relative deviation of the measured cyclotron frequency ratios from the calculated ratios $\varepsilon(r)/r$ is plotted as a function of the mass difference $\Delta m = (m_{ref} - m_{ioi})$ between the reference ion and the ion of interest in Fig.~\ref{fig:clusters-ratios}. By fitting the data with a straight line, which is forced to pass through the origin, a mass-dependent shift 
\begin{equation} \label{eq:mass-error}
\frac{\delta_m r}{r} = -2.35(81) \times 10^{-10} / \textnormal{u} \times (m_{ref} - m_{ioi})
\end{equation}
was obtained.

The cyclotron frequency ratios were corrected for the obtained mass-dependent effect and the reduced chi-square $\chi ^2 / N$ for the $(r_{corr}-r_{calc})$ was greater than one, indicating the presence of an additional residual uncertainty. The residual uncertainty of
\begin{equation} \label{eq:residual-error}
\frac{\delta_{res} r}{r} = 9 \times 10^{-9}
\end{equation}
was quadratically added to the frequency ratios to satisfy the condition $\chi ^2 / N \leq 1$. The relative deviation of the corrected cyclotron frequency ratios with the included residual uncertainty is shown in Fig.~\ref{fig:clusters-ratios-corr}.

If only the carbon cluster measurements are taken into the analysis, the similar systematic uncertainties are obtained:  $\delta_m r / r = -2.39(87) \times 10^{-10} / \textnormal{u} \times \Delta m$ and $\delta_{res} r / r = 9.6 \times 10^{-9}$. If all the data are restricted to $|m_{ref} - m_{ioi}| \leq 12$~u, the mass-dependent shift and residual uncertainty are $-2.3(21) \times 10^{-10} / \textnormal{u} \times \Delta m$ and $5.3 \times 10^{-9}$, respectively. The previous cross-reference mass measurements with carbon cluster ions performed at JYFLTRAP using the TOF-ICR method with Ramsey excitation patterns resulted in the systematic uncertainties $\delta_m r / r = -7.8(3) \times 10^{-10} / \textnormal{u} \times \Delta m$ and $\delta_{res} r / r = 1.2 \times 10^{-8}$ for the data with $|m_{ref} - m_{ioi}| \leq 48$~u and $\delta_m r / r = -7.5(4) \times 10^{-10} / \textnormal{u} \times \Delta m$ and $\delta_{res} r / r = 7.9 \times 10^{-9}$ for the data with $|m_{ref} - m_{ioi}| \leq 24$~u \cite{Elomaa2009}.

The systematic uncertainties $\delta_m r / r$ and $\delta_{res} r / r$ impose a limit on the accuracy of mass determination at JYFLTRAP. However, it is worth noting, that in mass measurements with the mass doublets ($A_{ref} = A_{ioi}$) these systematic uncertainties are cancelled out \cite{Roux2013} and the accuracy level better than $10^{-9}$ \cite{Nesterenko2019} can be reached. Such accuracy can also be obtained in mass measurements with ions differing by $\Delta m = 2$~u \cite{Nesterenko2018,Nesterenko2020}. In general, it is very rare that reference mass is more than 12u away from the ion-of-interest.

\section{Conclusion}

In this work the systematic uncertainties of the mass measurements with the PI-ICR technique at the JYFLTRAP setup are discussed. The uncertainties related to the distortion of the ion motion projection onto the detector can be significantly reduced and maintained at a negligible level compared to the statistical uncertainties in the measurements with small angles $\alpha_c$ between the magnetron and the cyclotron phase spots. The systematic uncertainty due to the fluctuations of the magnetic field is $\delta B / (B \delta B) = 2.01(25) \times 10^{-12} \textnormal{ min}^{-1}$, which is suitable for long-term measurements. A 5-hour measurement between two reference measurements would introduce a relative uncertainty of $6.03(75) \times 10^{-10}$, which is much smaller than a typical statistical uncertainty obtained for weakly produced exotic nuclides. The effect of the collisions of the stored ions with residual gas in the measurement trap results in a significant damping of the cyclotron motion and increase in the size of the ion distribution of the cyclotron phase spot with increasing the phase accumulation time $t_{acc}$. It limits the resolution and accuracy of the angle determination and, therefore, the cyclotron frequency determination. The phase accumulation times at JYFLTRAP are typically chosen up to 1.2 s. The damping effect is more pronounced for the carbon-cluster ions than for the monoatomic ions with the similar masses due to different ion mobilities.

In the case when more than one ion species is simultaneously stored in the measurement trap the accumulated magnetron phase position has to be corrected due to finite durations of the excitation pulses. Derivation of an analytic expression of the phase evolution of radial ion motion during the application of rf fields made it possible to accurately correct for this effect.
Also, a significant effect of the ion-ion interactions between different ion species in the measurement trap was observed in the count-rate class analysis. To take this effect into account more accurately the combined efficiency of the MCP detector and the data acquisition system was measured as a function of number of detected ions.
Typically, the count-rate effect is not observed for a small number of detected ions ($\sim 1-5$ ions/bunch) of the same ion species at JYFLTRAP, but with different ion species present, the effect was found to be significant.

The cross-reference mass measurements with the carbon-cluster ions allowed to determine the systematic uncertainties in the case when the ion of interest and the reference ion are not a mass doublet. The mass-dependent and residual uncertainties are $\delta_m r/r = -2.35(81) \times 10^{-10} / \textnormal{u} \times \Delta m$ and $\delta_{res} r/r = 9 \times 10^{-9}$, respectively,
for the data with $|\Delta m| = |m_{ref} - m_{ioi}| \leq 36$~u and $\delta_m r/r = -2.3(21) \times 10^{-10} / \textnormal{u} \times \Delta m$ and $\delta_{res} r/r = 5.3 \times 10^{-9}$, respectively,
for the data with $|\Delta m| \leq 12$~u.
For example, mass measurement of ions with $\Delta m = 20$~u in the mass region of $A \sim 100$ the mass-dependent and residual uncertainties introduce a systematic uncertainty of about 1~keV/$c^2$. This level of accuracy is enough for the mass values of radioactive nuclides far from stability needed for astrophysics or nuclear structure studies. For more precise mass measurement at the level of $10^{-9}$ and better, the ion of interest and reference should be within $|\Delta m|\le 2$ u \cite{Nesterenko2020} or ideally $A/q$ doublets. The mass measurements with the carbon-cluster ions are in-line with the earlier measurements \cite{Elomaa2009}. It is also worth noting that in most mass measurements at JYFLTRAP the mass difference between the reference ion and the ion of interest is $|\Delta m| \leq 12$~u, allowing $5.3\times10^{-9}$ precision to be reached.

\section*{Acknowledgements}

This work has been supported by the European Union’s Horizon 2020 research and innovation program under grant agreement No.771036 (ERC CoG MAIDEN) and by the Academy of Finland under projects No.295207 and 327629.

\bibliography{mybibfile}

\end{document}